\documentclass[fleqn,usenatbib]{mnras}

\usepackage{mathptmx}
\usepackage[T1]{fontenc}
\usepackage{graphicx}	% Including figure files
\usepackage{amsmath}	% Advanced maths commands
\usepackage{amssymb}	% Extra maths symbols

\newcommand{\kev}{keV}
\newcommand{\ecut}{E_{\mathrm{cut}}}
\newcommand{\chandra}{\textit{Chandra}}
\newcommand{\integral}{\textit{INTEGRAL}}

\newcommand{\bat}{\textit{Swift}-BAT}

\newcommand{\nustar}{\textit{NuSTAR}}
\newcommand{\xmm}{\textit{XMM-Newton}}

\newcommand{\fe}{Fe~K$\alpha$}
\newcommand{\etal}{et al.}
\DeclareMathOperator{\str}{str}
\title[An Evolving AGN Reflection Fraction with $z$]{The evolving X-ray spectrum of active galactic nuclei: evidence for an increasing reflection fraction with redshift}

\author[M. S. Avirett-Mackenzie \& D. R. Ballantyne]{
M. S. Avirett-Mackenzie\thanks{E-mail: msam7@gatech.edu}
and D. R. Ballantyne
\\
Center for Relativistic Astrophysics, School of Physics, Georgia
  Institute of Technology, 837 State Street, Atlanta, GA 30332-0430, USA\\
}

\date{Accepted XXX. Received YYY; in original form ZZZ}

\pubyear{2018}

\begin{document}
\label{firstpage}
\pagerange{\pageref{firstpage}--\pageref{lastpage}}
\maketitle

% Abstract of the paper
\begin{abstract}
The cosmic X-ray background (XRB) spectrum and active galaxy number
counts encode essential information about the spectral evolution of
active galactic nuclei (AGNs) and have been successfully modeled by
XRB synthesis models for many years. Recent measurements of the
$8$--$24$~\kev\ AGN number counts by \nustar\ and \bat\ are unable to
be simultaneously described by existing XRB synthesis models,
indicating a fundamental breakdown in our understanding of AGN
evolution. Here we show that the $8$--$24$~\kev\ AGN number counts can
be successfully modeled with an XRB synthesis model in which the
reflection strength ($R$) in the spectral model increases with
redshift. We show that an increase in reflection strength with
redshift is a natural outcome of (1) connecting $R$ to the incidence
of high column density gas around AGNs, and (2) the growth in the AGN
obscured fraction to higher redshifts. In addition to the redshift
evolution of $R$, we also find tentative evidence that an increasing
Compton-thick fraction with $z$ is necessary to describe the
$8$--$24$~\kev\ AGN number counts determined by \nustar. These results
show that, in contrast to the traditional orientation-based AGN
unification model, the nature and covering factor of the absorbing gas
and dust around AGNs evolve over time and must be connected to
physical processes in the AGN environment. Future XRB synthesis models
must allow for the redshift evolution of AGN spectral parameters. This approach
may reveal fundamental connections between AGNs and their host galaxies.
\end{abstract}

% Select between one and six entries from the list of approved keywords.
% Don't make up new ones.

%%%%%%%%%%%%%%%%%%%%%%%%%%%%%%%%%%%%%%%%%%%%%%%%%%

%%%%%%%%%%%%%%%%% BODY OF PAPER %%%%%%%%%%%%%%%%%%

\section{Introduction}
\label{sect:intro}
At rest frame energies above 2~\kev, the X-ray spectral energy
distribution (SED) of unabsorbed active galactic nuclei (AGNs) can be
described as the sum of just two interconnected components. The first
of these is a cutoff power-law spectrum with photon index $\Gamma$ and
cutoff energy $\ecut$ that is consistent with being produced by the
Compton up-scattering of ultraviolet (UV) photons from the accretion
disc in a hot tenuous corona
\citep*[e.g.,][]{grv79,hm91,hm93,hmg94}. The rapid variability of the
power-law, as well as novel micro-lensing size measurements, both
indicate that the corona is compact and located close to the central
supermassive black hole (SMBH; e.g.,
\citealt{rm13,zog13,mac15,kara16}). The second component of an AGN
spectrum is one or more reflection spectra produced by the interaction
of the power-law with cold dense gas situated out of the line-of-sight
\citep*[e.g.,][]{gf91,matt91,rf93,rfy99,rf05,gk10}. At energies above
2~\kev, the reflection spectrum adds an \fe\ line at $6$--$7$~\kev\ and
a Compton `hump' between $\approx 20$--$40$~\kev\ to the intrinsic
power-law. Deep exposures of bright nearby AGNs find a low-contrast,
relativistically broadened \fe\ line along with a strong and narrow
line `core' at 6.4~\kev\ indicating that both the inner accretion disc
and distant optically thick gas in the AGN environment contribute to
the reflection signal
\citep*[e.g.,][]{nan07,fero10,bn11,pat12,walton13,mnp16}. Typically
only the narrow \fe\ line is measurable in fainter AGNs outside the
local Universe \citep[e.g.,][]{main07,march16}.  

Efficient X-ray reflection only occurs in relatively cold
Compton-thick (or near Compton-thick) gas
\citep[e.g.,][]{gf91,matt91}. Since even unobscured AGNs exhibit this
distant reflection signature, the gas must lie out of the
line-of-sight, but still subtend a significant solid-angle as seen
from the X-ray source. Identifying the origin of the distant reflector
has been a significant observational challenge as, until recently, the
narrow \fe\ line was the only direct probe of this
spectrum. Nevertheless, measurements of the \fe\ line width from
\chandra\ observations indicated that the reflector is likely
connected to the obscuring gas associated with AGN unification models
\citep*[e.g.,][]{nan06,shu10,shu11}, and the equivalent width of the
\fe\ line appears to decrease with AGN luminosity (an `X-ray Baldwin
Effect'; e.g., \citealt{it93,shu12,ricci13,ricci14,march16}) similar
to the observed decrease in the fraction of obscured, or Type 2, AGNs
($f_2$) with luminosity
\citep[e.g.,][]{has08,burlon11,merloni14,georg17}. Therefore, the
distant reflector appears to probe the high column density regions of
the obscuring gas around AGNs. Since directly detecting AGNs that are
absorbed by Compton-thick gas is extremely challenging at all
wavelengths \citep{ha18}, the distant reflector can be used as a direct
probe of the Compton-thick gas in the AGN environment. Crucially, if
reflection signatures can be measured in AGNs at higher redshifts,
then they could be used to probe the evolution of Compton-thick gas
over time, which will be important for models of AGN fueling and
feedback \citep[e.g.,][]{hopk06}. 

The strength of the distant reflector is also important for fitting
the shape of the cosmic X-ray background (XRB). XRB synthesis models
rely on a model AGN SED that is then integrated
over luminosity and redshift
\citep*[e.g.,][]{com95,tu05,bem06,gch07,db09,tuv09,ball11,ueda14}. It was established early on \citep[e.g.,][]{ueda03} that a significant
reflection strength (denoted by $R$) is needed to account for the
observed peak of the XRB at $\approx 30$~\kev. As the XRB itself is a
poor constraint on $R$ \citep{aky12}, different XRB models employ
different assumptions on the value of $R$, but most assumed a constant
$R \sim 1$, a value based on observations of bright, nearby AGNs
\citep[e.g.,][]{ricci17} and corresponds to an isotropic source above an infinite disk. Some models included the decrease of $R$
with luminosity \citep[e.g.,][]{gch07,db09}, but all assumed that
there was no redshift dependence. Despite the simplicity of the
assumptions on $R$, the XRB models could successfully account for the
XRB spectrum and much of the $\la 10$~\kev\ data
\citep[e.g.,][]{ball11}. However, a hint that these models were
incomplete was noted by \citet{ajello12} who found that none of the
available XRB models could describe the \bat\ $15$--$55$~\kev\ AGN number
counts. 

The broad ($3$--$79$~\kev) bandpass and focusing optics provided by
\nustar\ \citep{harr13} now allow the full AGN reflection spectrum
(i.e., \fe\ line and Compton hump) to be analyzed for AGNs far beyond
the local Universe. Most of the high-$z$ AGNs detected by \nustar\ in
deep extragalactic survey fields \citep{civano15,mull15,lans17} are
still too faint to perform spectral fitting, but \citet{zappa18} was
able to estimate $R$ from 63 of the brightest sources with a median
$z=0.58$ and found a significant decrease of $R$ with X-ray
luminosity. Similar results were also found by \citet{delmoro17} who
analyzed stacked \nustar\ spectra from the deep fields. In addition,
both studies found that the typical values of $R$ were $\la
1$. Unfortunately, neither group was able to disentangle redshift
effects from the flux limited samples provided by the survey
fields. These results are in striking contrast with the integrated
results from the \nustar\ surveys. When both the
$8$--$24$~\kev\ luminosity function and numbers counts are computed
from the \nustar\ data, large ($R \sim 2$) values of the reflection
strength are needed in the XRB models to fit these data
\citep{aird15,harr16}. Even more interesting is the fact that, as was
anticipated by \citet{ajello12}, these high $R$ models cannot
simultaneously fit both the \nustar\ and \bat\ number counts
\citep{harr16}. This problem was recently discussed by \citet{ag19}
who confirmed the fundamental shape difference in the survey data
produced by the two missions, but were unable to identify any plausible systematic
statistical or instrumental mechanism for the disagreement. As the \bat\ counts are predominantly from AGNs at $z
< 0.1$, and the median redshift of the \nustar\ catalog is $z=0.76$
\citep{harr16}, the inability for models to explain both data sets may
be indicating the presence of redshift evolution in the average AGN
spectrum. Since the $8$--$24$~\kev\ band probed by \nustar\ and \bat\ is sensitive to the Compton
hump in the reflection spectrum, an evolving $R$ is a prime candidate
to explain the discrepancy, and would indicate an increase in
Compton-thick gas in the AGN environment at higher redshifts.

This paper presents a new X-ray background synthesis model that can
evolve the average spectral parameters of AGNs in both luminosity and
redshift with an initial focus of developing a model that can explain
the discrepancy between the $8$--$24$~\kev\ \nustar\ and \bat\ number
counts. The next section describes the details of the model and its
underlying assumptions, and we begin our analysis in
Sect.~\ref{sect:toymodel} by presenting the $8$--$24$~\kev\ number
counts problem and considering various solutions. Based on the results
of this preliminary work, Sect.~\ref{sect:nhrefl} describes an XRB
model with a physically-motivated evolving $R$ and compares it to the
$8$--$24$~\kev\ \nustar\ and \bat\ number counts. The results are
discussed in Sect.~\ref{sect:discuss} and conclusions are presented in
Sect.~\ref{sect:concl}. The paper assumes a standard flat $\Lambda$CDM
cosmology: $H_0=70$~km~s$^{-1}$~Mpc$^{-1}$, $\Omega_m=0.3$, and
$\Omega_{\Lambda}=0.7$. 

% Examine description from Ballantyne, 2011
% Discuss the way the model is built up, choices of R(L) model
% Mention Gaussian average of gamma
% Talk about luminosity function (e.g. we used Ueda 2014)
\section{Description of the X-ray Background Synthesis Model}
\label{sect:model}
% MSAM to draft. DRB to fill in details.
An XRB synthesis model consists of computing both the spectrum of the
XRB as a function of energy ($E$, in \kev), 
\begin{equation}
\begin{split}
I(E) = \frac{c}{H_0} &\int_{z_{\min}}^{z_{\max}} \int_{\log L_{\mathrm{X}}^{\min}}^{\log L_{\mathrm{X}}^{\max}} \frac{d \Phi(L_{\mathrm{X}}, z)}{d \log L_{\mathrm{X}}} \\ 
&\times \frac{S_E(L_{\mathrm{X}}, z) d_l^2}{(1 + z)^2(\Omega_m(1 + z)^3 + \Omega_{\Lambda})^{1/2}} d \log L_{\mathrm{X}} dz,
\end{split}
\label{eq:xrb}
\end{equation}
and the differential AGN number counts as a function of flux ($S$,
defined over some energy band), 
\begin{equation}
\begin{split}
\frac{dN}{dS}(S) = \frac{K_{\str}^{\deg} c}{\ln 10 H_0} &\int_{z_{\min}}^{z_{\max}} \frac{d \Phi(L_{\mathrm{X}}, z)}{d \log L_{\mathrm{X}}} \\
&\times \frac{d_l^2}{(1 + z)^2(\Omega_m(1 + z)^3 + \Omega_{\Lambda})^{1/2}} \frac{dz}{S}.
\end{split}
\label{eq:counts}
\end{equation}
In these two equations $d \Phi(L_{\mathrm{X}}, z)/d \log
L_{\mathrm{X}}$ denotes the hard X-ray luminosity function (HXLF),
$S_E(L_{\mathrm{X}}, z)$ is an average absorbed AGN X-ray spectrum
with an unabsorbed $2$--$10$~\kev\ luminosity $L_{\mathrm{X}}$ at
redshift $z$, $d_l$ is the luminosity distance to $z$, and
$K_{\str}^{\deg} = 3 \times 10^4$ converts the number counts from
$\str^{-1}$ to $\deg^{-2}$. The integrals are evaluated from $z_{\min}
= 0$ to $z_{\max} = 6$ and $\log
(L_{\mathrm{X}}^{\min}/\mathrm{erg\ s^{-1}}) = 41.5$ to $\log
(L_{\mathrm{X}}^{\max}/\mathrm{erg\ s^{-1}}) = 48$. The HXLF described
by \citet{ueda14} is used in all calculations.  

The unabsorbed AGN spectrum $S_E$ is constructed from a series of
cutoff power-law spectra modified by reflection that are computed
using the \textsl{pexmon} model \citep{nan07} provided in
\textsc{xspec} v.12.9.1 \citep{arn96}. All \textsl{pexmon} spectra
assume Solar abundances and an inclination angle of 60
degrees. Following \citet{gch07}, $S_E$ is computed by Gaussian
averaging 11 individual spectra (all with the same
$R$ and $E_{\mathrm{cut}}$) around a central photon index of $\langle
\Gamma \rangle = 1.9$ \citep[e.g.,][]{zappa18} with $\sigma_{\Gamma} =
0.3$ \citep[e.g.,][]{march16}. The cutoff energy is fixed at
$E_{\mathrm{cut}}=220$~\kev, consistent with recent measurements by
\nustar\ and \bat\ \citep[e.g.,][]{ricci18,tort18}, as well as the
average $E_{\mathrm{cut}}$ found by \citet{ball14} at $z\approx
0$.  The
redshift and luminosity dependence of $R$ is investigated in two different ways which
are described in Sects.~\ref{sect:toymodel} and~\ref{sect:nhrefl}.

Once the unabsorbed $S_E$ is defined at a specific $L_{\mathrm{X}}$
and $z$ it is subject to soft X-ray absorption by a column density
$N_{\mathrm{H}}$ local to the AGN. Absorbed $S_E$ are calculated for
each $\log (N_{\mathrm{H}}/\mathrm{cm}^{-2})=20, 20.5,\ldots, 24.5,
25$ using the \citet{mcm83} photoelectric cross-sections for
Compton-thin gas (e.g., $\log (N_{\mathrm{H}}/\mathrm{cm}^{-2}) \leq
23$). The transmitted spectra through larger column densities is
significantly affected by Compton scattering and is calculated using
suppression factors determined by \textsc{mytorus} \citep{yaq12}. A
scattered pure reflection spectrum, with a scattering fraction of
$2$\%, calculated following the same procedure as above (i.e., the
Gaussian average over $\Gamma$) is set equal to the $\log
(N_{\mathrm{H}}/\mathrm{cm}^{-2})=25$ spectrum and is added to the
$\log (N_{\mathrm{H}}/\mathrm{cm}^{-2})=24.5$ spectrum. The final
$S_E(L_{\mathrm{X}},z)$ is constructed by summing the individual
absorbed spectra while weighting for the fraction of obscured AGNs,
$f_2$ (defined as the fraction of AGNs obscured by $\log
(N_{\mathrm{H}}/\mathrm{cm}^{-2}) \geq 22$), and the \citet{burlon11}
$N_{\mathrm{H}}$ distribution. The obscured AGN fraction is observed
to be a function of both X-ray luminosity and redshift
\citep[e.g.][]{ueda03,laf05,bem06,has08,merloni14,ueda14,liu17}. We
follow the \citet{burlon11} description of the luminosity dependence
at $z=0$, but include a redshift dependence for AGNs with $\log
(L_{\mathrm{X}}/\mathrm{erg\ s^{-1}}) \gtrsim 43.5$ where the evidence
for redshift evolution is the strongest
\citep[e.g.,][]{merloni14,liu17}. To be conservative the redshift
dependence is halted at $z=2$; therefore, $f_2$ is determined by 
\begin{equation}
f_2(L_{\mathrm{X}},z) = \left\{ \begin{array}{l}
0.8 e^{-L_{\mathrm{X}}/{L_c}} + 0.2(1 + z)^{\xi} \left(1 - e^{-L_{\mathrm{X}}/L_c} \right),\ \ z \leq 2 \\
0.8 e^{-L_{\mathrm{X}}/{L_c}} + 0.2(1 + 2)^{\xi} \left(1 - e^{-L_{\mathrm{X}}/L_c} \right),\ \ z > 2 \end{array}\right.
\label{eq:type2}
\end{equation}
where $\log (L_c/\mathrm{erg\ s^{-1}})=43.7$ and $\xi=0.48$ \citep{ueda14}. 
% and $\xi=0.48$ \citep{ueda14}.
% Discuss choice of \xi and justification -- 0.48 for toy model, 1.3 for N_H

Compton-thick AGNs, defined as AGNs obscured by column densities $\log
(N_{\mathrm{H}}/\mathrm{cm}^{-2}) \geq 24$, are extremely faint in the
$2$--$10$~\kev\ band and are not included in the \citet{ueda14}
HXLF. In fact, despite significant effort, the fraction of AGNs that
are Compton-thick is highly uncertain beyond the local Universe
\citep[e.g.,][]{burlon11,buch15,ricci15,lanz18,masini18}. Synthesis
models have shown that a significant population of Compton-thick AGNs
is needed in order to fit the peak of the XRB spectrum
\citep[e.g.,][]{ueda03,gch07,db09,tuv09,ball11}. Unfortunately, the
fraction of Compton-thick AGNs needed in the synthesis models is
degenerate with the assumed HXLF \citep{db09}, as well as the
reflection fraction and high-energy cutoff of the assumed AGN spectrum
\citep{aky12}. In addition, there is evidence that Compton-thick AGNs
may be more common in high accretion rate episodes and therefore this
population evolves differently than less obscured AGNs
\citep{db10,dale15}. We start by following the traditional approach of
assuming a fixed Compton-thick fraction, defined so that the $z=0$
space density of Compton-thick AGNs with $\log
(L_{\mathrm{X}}/\mathrm{erg\ s}^{-1}) \geq
43.2$ is $4\times 10^{-6}$~Mpc$^{-3}$ \citep{buch15}. This corresponds
to a fraction of $9\times 10^{-3}$ at a $2$--$10$~\kev\ flux of
$10^{-14}$~erg~cm$^{-2}$~s$^{-1}$ (compare to Fig.\ 6 of
\citealt{lanz18}) and a fraction of $0.06$ at a $8$--$24$~\kev\ flux
of $2.7\times 10^{-14}$~erg~cm$^{-2}$~s$^{-1}$ (compare to Fig.\ 15 of \citealt{masini18}), where the fraction is defined as
the ratio of the number of AGNs with $\log
(N_{\mathrm{H}}/\mathrm{cm}^{-2}) \geq 24$ to the number of all AGNs
(including those that are Compton-thick). Both of these fractions are
approximately half the observed values at these fluxes, suggesting
a change in the Compton-thick fraction beyond the local Universe. The Compton-thick AGNs are
distributed equally over the $N_{\mathrm{H}}$ bins $\log
(N_{\mathrm{H}}/\mathrm{cm}^{-2}) = 24,\ 24.5,\ 25$. 

% Describing our proof of concept with (1 + z)^{\alpha}
% Includes results
\section{Evidence for an Evolving Reflection Fraction}
\label{sect:toymodel}
\begin{figure*}
\includegraphics[angle=-90,width=0.48\textwidth]{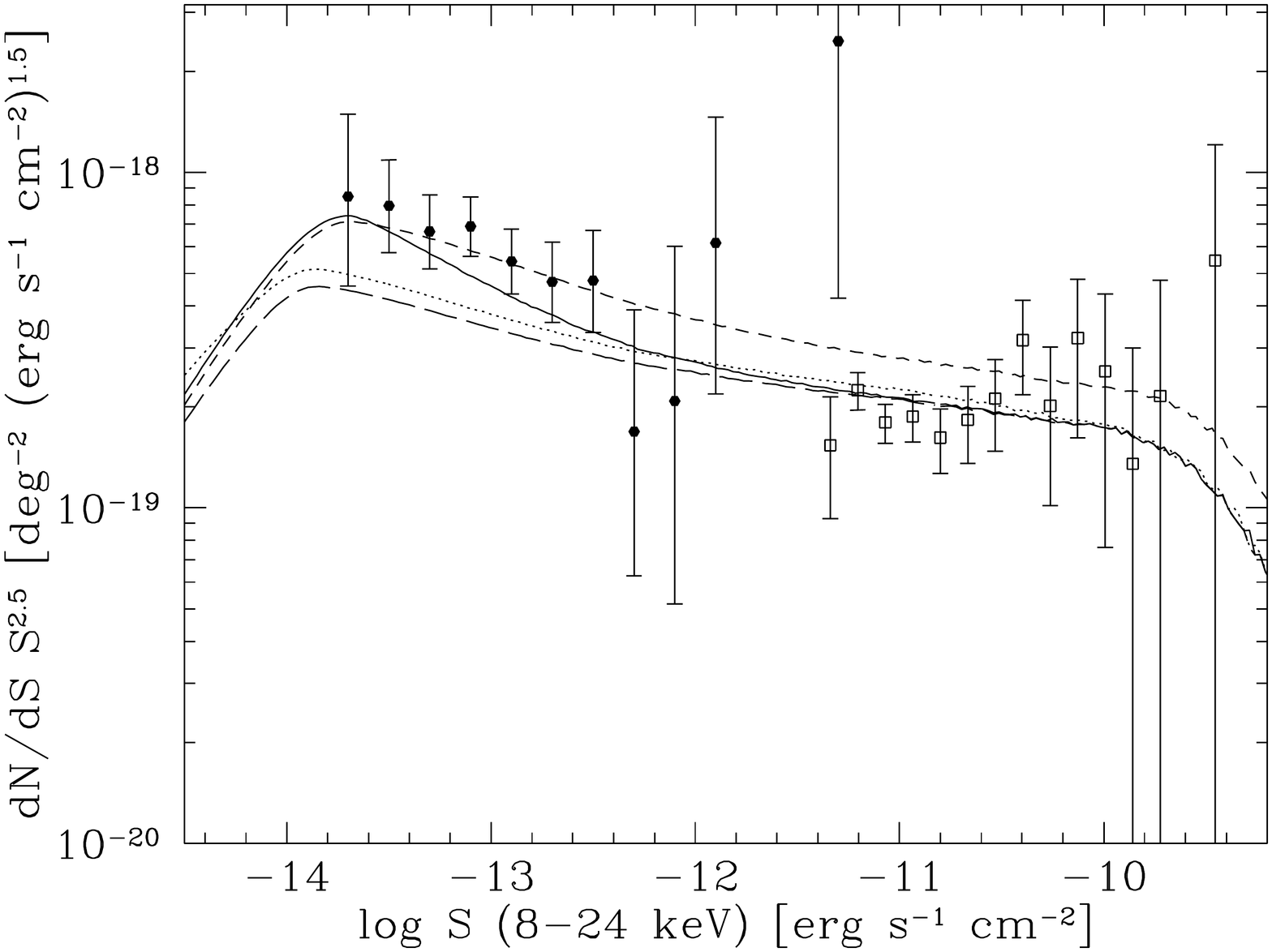}
\includegraphics[angle=-90,width=0.48\textwidth]{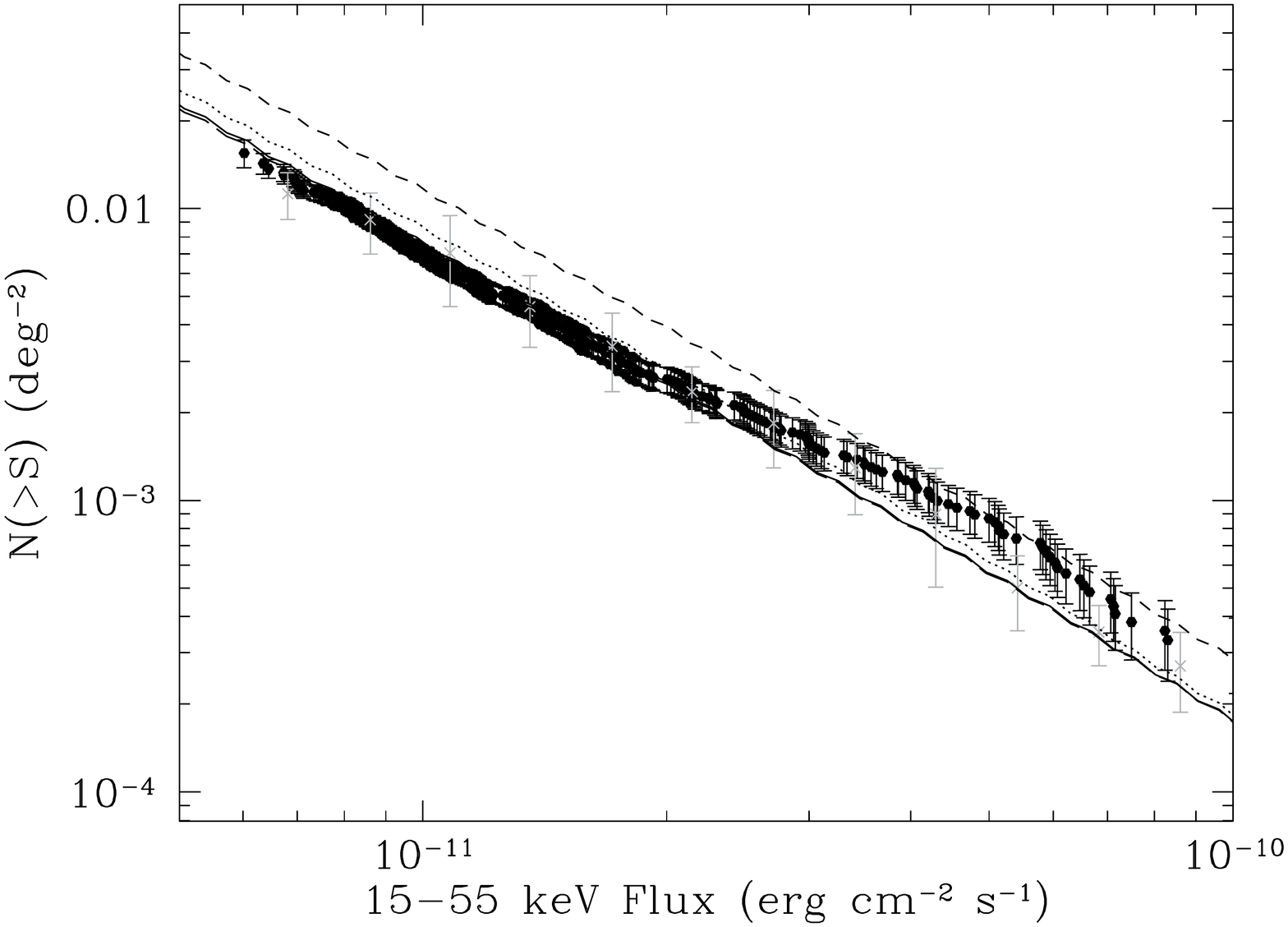}
\caption{\textbf{(Left)} The differential AGN $8$--$24$~\kev\ number
  counts compared with four XRB models described in
  Sect.~\ref{sect:toymodel}. The short-dashed line plots a traditional
  XRB model in which the reflection strength is fixed at $R=1.2$
  independent of both $L_{\mathrm{X}}$ and $z$. The long-dashed line
  shows the XRB spectrum predicted when $R$ has the luminosity
  dependence described by eq.~\ref{eq:Ldepend} with $a=-0.26$ and
  $b=10.91$. The solid line plots the XRB spectrum when $R$ is a
  function of both $L_{\mathrm{X}}$ and $z$ (eq.~\ref{eq:toymodel})
  with $a=-0.26$, $b=10.91$, and $\alpha=2$. The filled data points
  are the \citet{harr16} \nustar\ measurements, and the squares are
  the \citet{ajello12} \bat\ measurements converted to the
  $8$--$24$~\kev\ band. It is clear that the model with a fixed
  $R=1.2$ can describe the \nustar\ data but not the \bat\ data, while
  the model with a luminosity-dependent $R$ (eq.~\ref{eq:Ldepend})
  that has significantly smaller values of $R$ can fit the \bat\ data
  but not the fainter \nustar\ measurements. The solid line which
  allows $R$ to grow with $z$ (eq.~\ref{eq:toymodel} with $\alpha=2$)
  provides a reasonable description of both datasets. Finally, the
  dotted line plots the same model as the one denoted by the
  long-dashed line, but with a Compton-thick fraction that is
  $2.76\times$ larger. This model is discussed in detail in
  Sect.~\ref{sub:comptonthick}. \textbf{(Right)} As a sanity check,
  the integrated $15$--$55$~\kev\ number counts are computed from the
  four XRB models and compared directly with the \bat\ data from
  \citet{ajello12} (densely plotted solid points). In addition, the
  AGN $17$--$60$~\kev\ number counts from \integral\ \citep{kriv10}
  are plotted as the grey points. As expected from the other panel,
  the XRB model with a fixed $R=1.2$ overpredicts most of these data,
  while the two evolving-$R$ models with the smaller Compton-thick
  fraction successfully account for the \bat\ and \integral\ number
  counts.} 
\label{fig:counts}
\end{figure*}

As described in Sect.~\ref{sect:intro}, previous XRB synthesis models
have been unable to simultaneously account for both the \bat\ and
\nustar\ number counts. To illustrate the magnitude and importance of
this problem, consider the left-hand panel of Figure~\ref{fig:counts}
which plots the differential $8$--$24$~\kev\ number counts from
\nustar\ and \bat.
As found by \citet{aird15} and \citet{harr16}, a XRB model with strong
reflection is needed to adequately fit the \nustar\ data. This
scenario is illustrated in the figure with the short-dashed line which
was computed using the model described in Sect.~\ref{sect:model}
assuming a fixed $R=1.2$ for all $L_{\mathrm{X}}$ and $z$. This model
provides a good description of the \nustar\ counts, but significantly
overestimates the \bat\ counts at higher fluxes. The right-hand panel
of Fig.~\ref{fig:counts} confirms this conclusion by directly
comparing the model predictions to the integrated $N(>S)$
$15$--$55$~\kev\ counts from \citet{ajello12}. 

\begin{figure}
\includegraphics[width=0.5\textwidth]{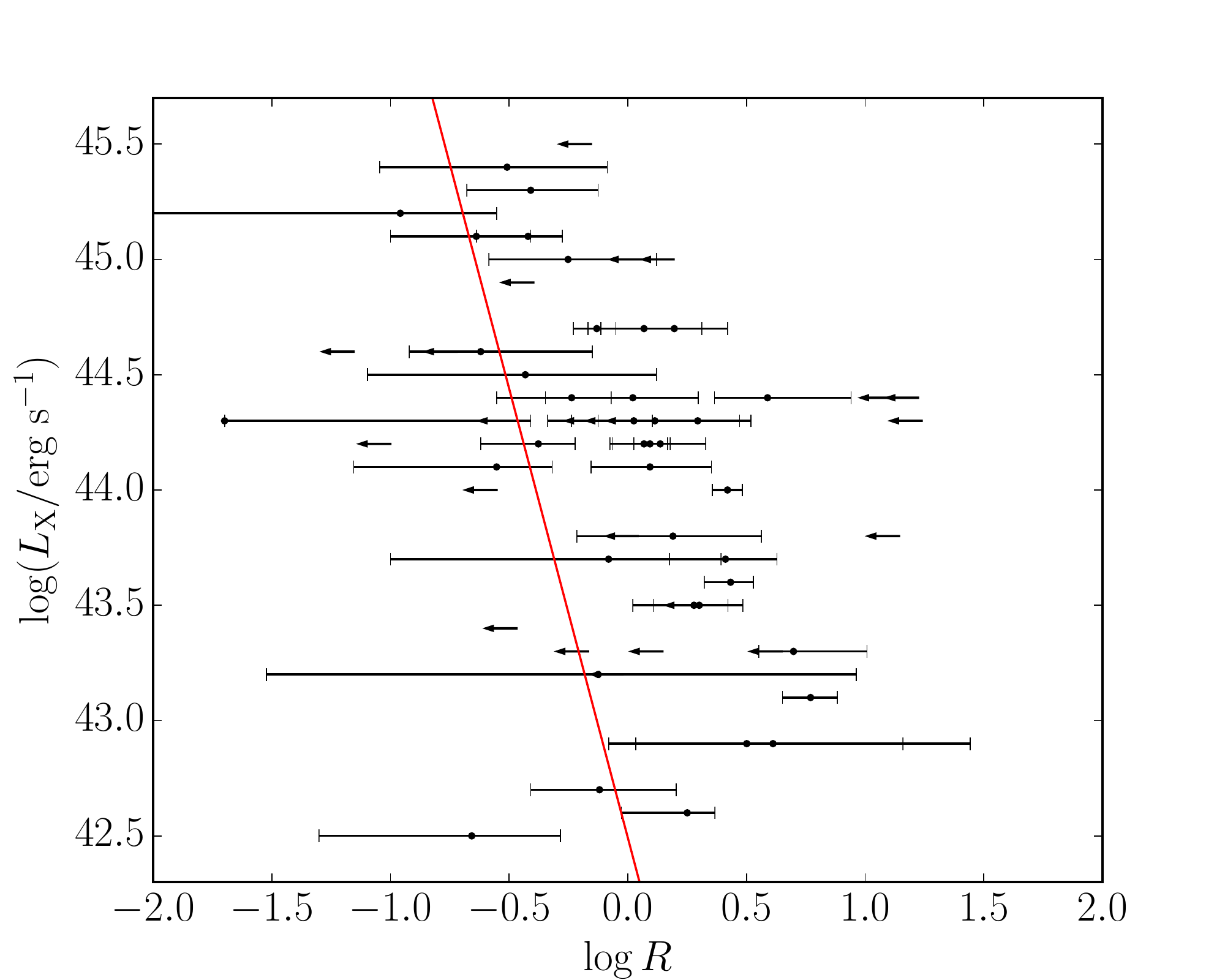}
\caption{Reflection parameter $R$ compared to unabsorbed luminosity
  $L_{\mathrm{X}}$ for the 63 AGNs presented by \citet{zappa18}. Error
  bars indicate sources whose value of $R$ is bounded from both above
  and below, while arrows indicate those whose $R$ is bounded only
  from above. The red line marks the fit of $\log R(L_{\mathrm{X}}) =
  -0.26\log L_{\mathrm{X}} + 10.91$.} 
\label{fig:lstsq}
\end{figure}

\begin{figure}
\includegraphics[angle=-90,width=0.5\textwidth]{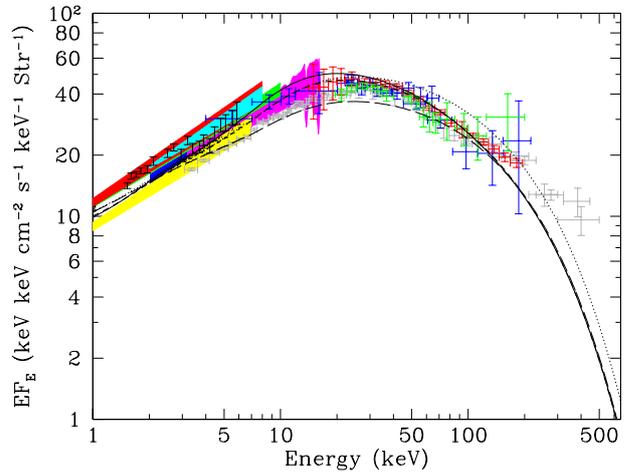}
\caption{The observed XRB spectrum is compared to four XRB synthesis
  models described in the text. The line styles denote the four models
  as in Fig.~\ref{fig:counts}. While both the short-dashed and solid
  lines provide reasonable fits to the observed XRB spectrum, the
  model with the consistently weak reflection described by
  eq.~\ref{eq:Ldepend} (long-dashed line) significantly underpredicts
  the data. The colored data and areas denote different measurements
  of the XRB spectrum: blue - \textit{ASCA} GIS \citep{kush02};
  magenta - \textit{RXTE} \citep{rev03}; green - \textit{XMM-Newton}
  \citep{lumb02}; red - \textit{BeppoSAX} \citep{vec99}; yellow -
  \textit{ASCA} SIS \citep{gen95}; cyan - \textit{XMM-Newton}
  \citep{dm04}; grey data - \textit{HEAO-1} \citep{gru99}; blue data -
  \textit{INTEGRAL} \citep{chur07}; red data - \textit{SWIFT} BAT
  \citep{ajello08}; black data - \textit{Swift}-XRT \citep{mor09};
  green data - \textit{INTEGRAL} \citep{tur10}.} 
\label{fig:cxrb}
\end{figure}

Current observations disfavour a model with a fixed $R$
\citep[e.g.,][]{delmoro17}, and instead suggest that $R$ is lower for
higher-luminosity AGNs \citep[e.g.,][]{zappa18}. To parameterize this
luminosity dependence we consider a simple power-law relation,
\begin{equation}
\label{eq:Ldepend}
\log R(L_{\mathrm{X}}) = a\log L_{\mathrm{X}} + b,
\end{equation}
where the $a$ and $b$ are determined by performing a least-squares fit
to the $L_{\mathrm{X}}$ and $R$ data provided by
\citet{zappa18}. Fitting to the central data points shown in
Figure~\ref{fig:lstsq} suggests the values $a = -0.29$; $b = 12.8$
while fitting to the lower and upper bounds on $\log R$ indicated in
the figure give the values $a = -0.26$; $b = 10.91$ and $a = -0.29$;
$b =  13.1$ respectively. However, evolutionary models using $a =
-0.29$ and $b \approx 13$ dramatically overestimate the XRB, while the
lower bound model predicts it within reasonable error. We thus chose
$a = -0.26$ and $b = 10.91$ as our luminosity evolutionary model. 
The long-dashed lines in Fig.~\ref{fig:counts} shows the predicted
$8$--$24$~\kev\ $dN/dS$ using this $R$--$L_{\mathrm{X}}$
 dependence. This XRB
model has uniformly lower values of $R$ than the previous one
($R=1.2$; short-dashed line) and cannot fit the
deep \nustar\ counts, but it does provide an excellent fit to the
\bat\ data. Interestingly, the $R$--$L_{\mathrm{X}}$ relationship was
determined by spectral fitting of individual \nustar\ observations
from the deep surveys (those selected to have $\log
S_{\mathrm{8-24\ keV}} \geq -13.15$; \citealt{zappa18}). Perhaps these
measurements of $R$ are more accurate than what is inferred by fitting
the \nustar\ counts at even fainter fluxes. However, examining the
predicted XRB spectrum from these two models (Fig.~\ref{fig:cxrb})
shows that the weaker reflection strengths predicted by
eq.~\ref{eq:Ldepend} significantly underpredicts the peak of the XRB,
while the fixed $R=1.2$ model provides a very reasonable fit to the
entire XRB spectrum. 
Therefore, the $R$--$L_{\mathrm{X}}$ model shown by the long-dashed
line must be missing additional reflection strength. Given the good
fit to the \bat\ data, this additional reflection strength must arise
at the fainter fluxes probed by the \nustar\ data which is dominated
by AGNs at much higher redshifts than the \bat\ sample. 

To test the idea of an increasing $R$ with $z$, we calculate a XRB
synthesis model with the following simple prescription for $R$, 
\begin{equation}
\label{eq:toymodel}
R(L_{\mathrm{X}},z) = \left\{ \begin{array}{l}
R(L_{\mathrm{X}})(1+z)^{\alpha},\ \ z \leq 2 \\
R(L_{\mathrm{X}})(1+2)^{\alpha},\ \ z > 2 
\end{array}\right.
\end{equation}
where $R(L_{\mathrm{X}})$ is given by eq.~\ref{eq:Ldepend} with
$a=-0.26$ and $b=10.91$, as before. The solid lines in
Figs.~\ref{fig:counts} and~\ref{fig:cxrb} show the result of this
calculation when $\alpha=2$. This toy model retains the successful fit
to the \bat\ counts, and provides an adequate description of both the
\nustar\ counts and the XRB spectrum. Thus, a reflection fraction that
increases with redshift appears to be a viable method to
simultaneously account for both the \nustar\ and
\bat\ $8$--$24$~\kev\ data.  

\subsection{The Impact of the Compton-thick Fraction}
\label{sub:comptonthick}
The three XRB models presented above have a Compton-thick fraction set by
measurements of the local space density of Compton-thick AGNs
\citep{buch15}. However, if this is not a representative value beyond
the local Universe, then its possible that simply increasing the
Compton-thick fraction may solve the tension between the \nustar\ and
\bat\ number counts. As the Compton-thick AGN SED peaks at similar
energies to the reflection hump \citep[e.g.,][]{aky12}, an increase in
the Compton-thick fraction will have a similar effect in the XRB model
as a larger reflection fraction. To test this possibility, a XRB model
was calculated using the $R(L_X)$ dependence of Eq.~\ref{eq:Ldepend},
but with a Compton-thick fraction $2.76\times$ larger than previously
used. This model is shown as the dotted lines in
Figs.~\ref{fig:counts} \&~\ref{fig:cxrb}. The figures show that the
larger Compton-thick fraction has a minor impact on the predicted
total $8$--$24$~\kev\ number counts. The reason for the modest effect
is that the Compton-thick AGN population is still a small component of
the total AGN population, comprising only $\approx 2$\% of all AGN at
a $2$--$10$~\kev\ flux of 10$^{-14}$~erg~cm$^{-2}$~s$^{-1}$ after
nearly tripling its contribution. Moreover, as seen by the XRB
spectrum (Fig.~\ref{fig:cxrb}), adding even more Compton-thick AGN is
not possible, as the enhanced Compton-thick model already skims
the top of the observed error-bars. The results of this experiment
clearly show that an increase in reflection fraction with $z$ is
necessary to resolve the mismatch between the \nustar\ and
\bat\ number counts. 

\begin{figure}
\includegraphics[angle=-90,width=0.48\textwidth]{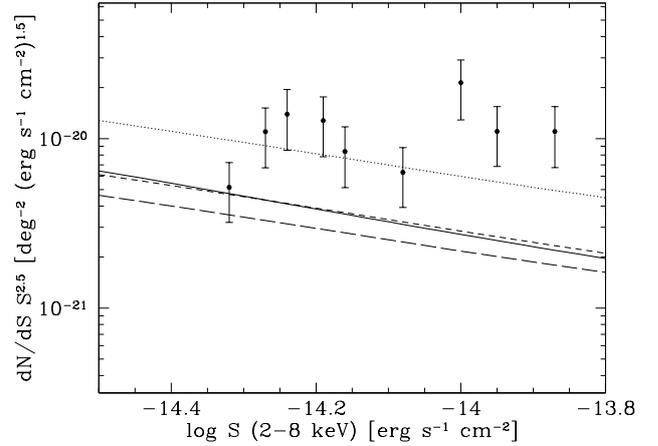}
\caption{\label{fig:comptonthick}The lines show the predicted
  Euclidean-normalized differential number counts of Compton-thick
  AGNs in the $2$--$8$~\kev\ band. The line styles denote the four
  different XRB models described in the text (see, e.g.,
  Fig.~\ref{fig:counts}). The data points are the observed
  Compton-thick counts in the $2$--$7$~keV energy band from the COSMOS
  Legacy survey \citep{ananna18,lanz18}. The very small correction
  between the $2$--$7$~keV and the $2$--$8$~\kev\ energy bands has
  been omitted for the purpose of this comparison.} 
\end{figure}

It is still of interest, however, to determine if the Compton-thick
fraction set by the local observed space-density is appropriate at the
fainter fluxes probed by \nustar. Figure~\ref{fig:comptonthick}
compares the Compton-thick AGN number counts predicted by the four XRB
models discussed here to deep measurements from the COSMOS Legacy
survey \citep{ananna18,lanz18}. 
The plot clearly shows that the three models with the
locally-calibrated Compton-thick fraction substantially underpredicts
the COSMOS data, including the two models that satisfactorily describe
the \nustar\ $8$--$24$~keV counts. In contrast, the model with the
enhanced Compton-thick fraction appears to more accurately describe
the Compton-thick counts at these fainter fluxes. This latter model
has a Compton-thick fraction of $0.14$ at a $8$--$24$~\kev\ flux
of $2.7\times 10^{-14}$~erg~cm$^{-2}$~s$^{-1}$, very similar to what
is measured from \nustar\ observations \citep{masini18}. The picture that
emerges from these experiments is one where both the reflection
fraction and possibly the Compton-thick fraction increase with $z$. In the next
section, we investigate this possibility further and explore a
potential physical explanation for these evolutions. 

% Describe R(N_H) model and present results
\section{A Physical Model for an Evolving Reflection Fraction}
\label{sect:nhrefl}

\subsection{Model Setup}
\label{sub:setup}
The previous section showed that a reflection fraction increasing with
$z$ can give a XRB model that satisfies both the \bat\ and
\nustar\ $8$--$24$~\kev\ differential number counts
(Fig.~\ref{fig:counts}). However, the values of the reflection
fraction predicted by this toy model grow to unreasonably large values
at high redshifts (e.g., $R=2.1$ at $z=1$ for $\log
(L_{\mathrm{X}}/\mathrm{erg\ s^{-1}})=43$, and this balloons to $R >
4$ by $z=2$). Thus, the na\"{\i}ve redshift evolution imposed in
Eq.~\ref{eq:toymodel} is clearly too extreme, and must be replaced by
a physically motivated method for increasing $R$ with $z$ that limits
the reflection fraction to more realistic values. 

As discussed in Sect.~\ref{sect:intro}, the reflection spectrum in
faint AGNs likely originates from reprocessing in the distant gas
responsible for AGN obscuration. Recently, \citet{llanz18} highlighted
the direct connection between $R$ and the obscuring gas in a sample of
\bat\ AGNs by finding a correlation between the reflected X-ray
luminosity (measured by \nustar) and the IR luminosity (measured by
\textit{WISE} and \textit{Herschel}). This result clearly indicates
that the X-ray, optical and UV radiation produced by the inner
accretion disk is reprocessed by a common structure. Since the
obscured fraction of AGNs, $f_2$, is observed to increase with $z$
\citep[e.g.,][]{merloni14,liu17}, then it is natural to expect that
the mean $R$ of AGNs will also increase with $z$. In addition, the
connection between $R$ and the obscuring gas has been made directly by
modeling the stacked spectra of Compton-thin AGNs detected by
\bat\ and \integral\ \citep{ricci11,vmg13,ew16}. In these studies,
stacked spectra separated into $N_{\mathrm{H}}$ bins were found to
have different values of $R$ with columns in the range $23 \leq
\log(N_{\mathrm{H}}/\mathrm{cm^{-2}}) \leq 24$ yielding the largest
reflection strengths. Such an effect would violate the simplest
unified AGN model where all values of $N_{\mathrm{H}}$ co-exist around
AGNs, independent of the line-of-sight obscuration. Instead, these
results imply that AGNs that are seen through moderate-to-heavy
amounts of obscuration exist in fundamentally different environments
than those observed through lower $N_{\mathrm{H}}$ columns, and are
therefore probing different AGN populations \citep{db10,buch15}. The
direct fitting of nearby \bat\ AGNs by \citet{llanz18} also found that
more obscured objects have larger $R$, although the analysis of
\citet{ricci17} gave the opposite conclusion which could be explained by modeling
degeneracies (see \citealt{llanz18}). 

\begin{figure}
\includegraphics[width=0.5\textwidth]{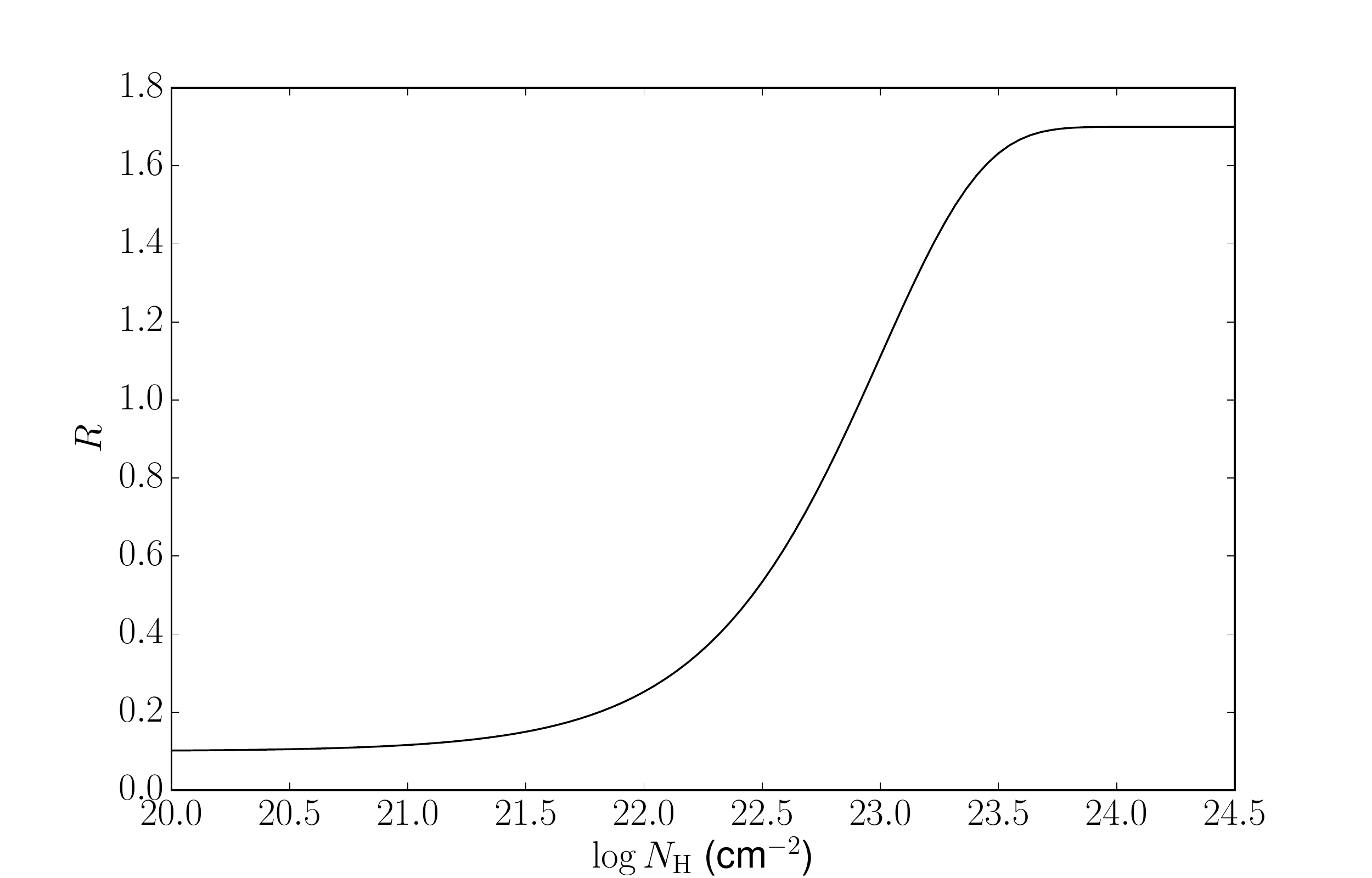}
\caption{An example of the relationship between reflection
  fraction, $R$, and the absorbing column density, $N_{\mathrm{H}}$
  (Eq.~\ref{eq:nhrefl}). The parameters $R_{\min}$ and $R_{\max}$ are
  fixed at $0.1$ and $1.7$, respectively, and the transition point
  $N_{\mathrm{H,mid}}$ is set to $10^{23}$~cm$^{-2}$.} 
\label{fig:rofnh}
\end{figure}

The combination of an increasing $f_2$ with $z$ and a correlation of $R$ with
$N_{\mathrm{H}}$ provides a physical template for the redshift
evolution in $R$ that is needed to fit the \nustar\ number
counts. From these two pieces of observational evidence a new model
for $R(L_{\mathrm{X}}, z)$ can be constructed which allows the
reflection fraction to increase to higher values of $N_{\mathrm{H}}$: 
\begin{equation}
R(N_{\mathrm{H}}[L_{\mathrm{X}},z])=R_{\min} e^{-N_{\mathrm{H}}/N_{\mathrm{H,mid}}} + R_{\max} \left(1 - e^{-N_{\mathrm{H}}/N_{\mathrm{H,mid}}} \right),
\label{eq:nhrefl}
\end{equation}
where $R_{\min}$ and $R_{\max}$ are the lower and upper bounds of $R$ and $N_{\mathrm{H,mid}}$ is the transition point between these values (Figure~\ref{fig:rofnh}).

Equation~\ref{eq:nhrefl} has implicit redshift and luminosity
dependencies in the following way. As redshift increases in the XRB
model, the fraction of obscured AGNs (i.e., those with
$\log(N_{\mathrm{H}}/\mathrm{cm}^{-2}) \geq 22$) rises as described by
eq.~\ref{eq:type2}. Therefore, the AGN spectra $S_E$ that are
constructed have a larger and larger proportion of heavily obscured
sources, which, as described by eq.~\ref{eq:nhrefl}, is associated
with a stronger reflection fraction. Therefore, the spectral model
used in the XRB calculation naturally has a larger $R$ at higher
redshifts. Likewise, the obscured fraction falls with luminosity in
Eq.~\ref{eq:type2}, and so the spectra of more luminous AGNs are
constructed with a larger fraction of weakly obscured AGNs which have
smaller $R$ (Fig.~\ref{fig:rofnh}). As a result, the
connection between reflection fraction and $N_{\mathrm{H}}$
(eq.~\ref{eq:nhrefl}) naturally describes similar redshift and
luminosity evolutions of $R$ as the toy models employed in
Sect.~\ref{sect:toymodel}.  

As seen in Fig.~\ref{fig:comptonthick}, a larger Compton-thick fraction at $\log
(S_{\mathrm{2-8\ keV}}/\mathrm{erg\ s^{-1}\ cm^{-2}}) \approx -14$ may
be necessary to match the COSMOS data implying that the evolution of
Compton-thick gas around AGNs is decoupled from that of the Compton-thin gas. To parameterize this effect in the XRB model, the Compton-thick fraction is allowed to increase with $z$ as a simple power-law,  
\begin{equation}
\label{eq:ct}
f_{\mathrm{CT}}(z) = \left\{ \begin{array}{l}
f_{\mathrm{CT,0}}(1+z)^{\gamma},\ \ z \leq 2 \\
f_{\mathrm{CT,0}}(1+2)^{\gamma},\ \ z > 2 
\end{array}\right.
\end{equation}
where $f_{\mathrm{CT,0}}$ is the local Compton-thick fraction used in
Sect.~\ref{sect:toymodel} and is determined by matching the measured
Compton-thick space density at $z \approx 0$ \citep{buch15}. 

\subsection{Results}
\label{sub:results}
In this section, we compare the results of this physically-motivated
XRB model to the \nustar\ and \bat\ number counts, the XRB spectrum,
and the COSMOS Compton-thick number counts. There are several
parameters in this new model (e.g., $R_{\mathrm{min}}$,
$R_{\mathrm{max}}$, $\gamma$), but as this is an exploratory model we
do not use a fitting method to determine the best-fit values of the
parameters. Rather, we explored the effects of each parameter
individually until we arrived at a result that best highlighted the
properties of the model. Therefore, the parameter values quoted here
should be viewed as a starting point from which to develop a more
sophisticated model of the evolution of the AGN environment and how it
impacts the observed X-ray spectra. 

\begin{figure}
\includegraphics[angle=-90,width=0.5\textwidth]{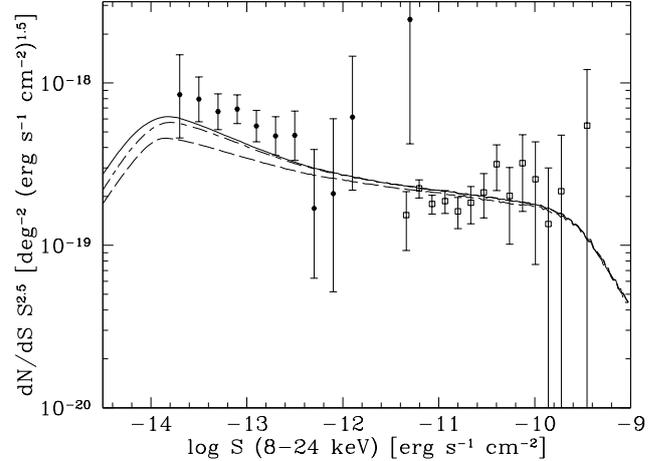}
\caption{The solid line shows the predicted Euclidean-normalized
  differential $8$--$24$~keV number counts from a $R(N_{\mathrm{H}})$
  model (eq.~\ref{eq:nhrefl}) with $R_{\mathrm{min}}=0.1$,
  $R_{\mathrm{max}}=1.7$, and $N_{\mathrm{H,mid}}= 10^{23}$~cm$^{-2}$. The model has a Compton-thick evolution
  parameter $\gamma=0.7$ and requires a more rapid increase in the
  Type 2 fraction, $\xi=1.3$ (eq.~\ref{eq:type2}) than the models
  discussed in Sect.~\ref{sect:toymodel}. A model with no
  Compton-thick evolution (i.e., $\gamma=0$) is shown as the
  short-dash-long-dash line and indicates that the $R(N_{\mathrm{H}})$
  model may not be enough on its own to account for the \nustar\ number
  counts. As a point of comparison, the dashed line plots the model
  from Fig.~\ref{fig:counts} that lacks any redshift evolution in $R$
  or the Compton-thick fraction. The data points are the same as
  Fig.~\ref{fig:counts}.} 
\label{fig:countsv2}
\end{figure}

We first consider a model with a fixed Compton-thick fraction (i.e.,
$\gamma=0$ in eq.~\ref{eq:ct}) in order
to isolate the effects of the $R(N_{\mathrm{H}})$ relationship
(eq.~\ref{eq:nhrefl}). The predicted $8$--$24$~\kev\ number counts for
a model with $R_{\mathrm{min}}=0.1$, $R_{\mathrm{max}}=1.7$ and
$N_{\mathrm{H,mid}}=10^{23}$~cm$^{-2}$ is shown as the
short-dash-long-dash-line in Figure~\ref{fig:countsv2}. The plot shows
that the increase in $R$ with $z$ driven by the changing $f_2$ does
indeed bend the number counts closer to the \nustar\ data, but falls just below the majority of the \nustar\ error-bars.  
At high fluxes the model is nearly identical to the one from
Fig.~\ref{fig:counts} that does not include any redshift evolution
(dashed line), but at fainter fluxes the larger fraction of obscured
AGNs with larger $R$ values causes the new model to diverge.
An interesting aspect of the model is that the factor describing the
increase in the Type 2 AGN fraction, $\xi$ (Eq.~\ref{eq:type2}),
was increased from $\xi=0.48$ (the value used in
Sect.~\ref{sect:toymodel}) to $\xi=1.3$ in order for the model
to lie close to the \nustar\ data. The higher value of $\xi$ ensured there
was a significant population of obscured AGNs at higher $z$ which
would have larger reflection fractions. Even larger values of $\xi$
were ruled out as those models produced an XRB $2$--$10$~\kev\ slope
much harder than observed ($\Gamma=1.45$; e.g., \citealt{capp17}). The
value $\xi=1.3$ is higher than what is commonly measured
\citep{has08,ueda14,liu17}, but is consistent with the results of
\citet{merloni14} who found similar values of $\xi$ in two out of
three luminosity bins. As pointed out by \citet{liu17} it is
challenging to accurately measure $\xi$ as the evolution may be
luminosity dependent (as found by \citealt{merloni14}) and appears to
weaken at $z > 2$. 

\begin{figure}
\includegraphics[width=0.5\textwidth]{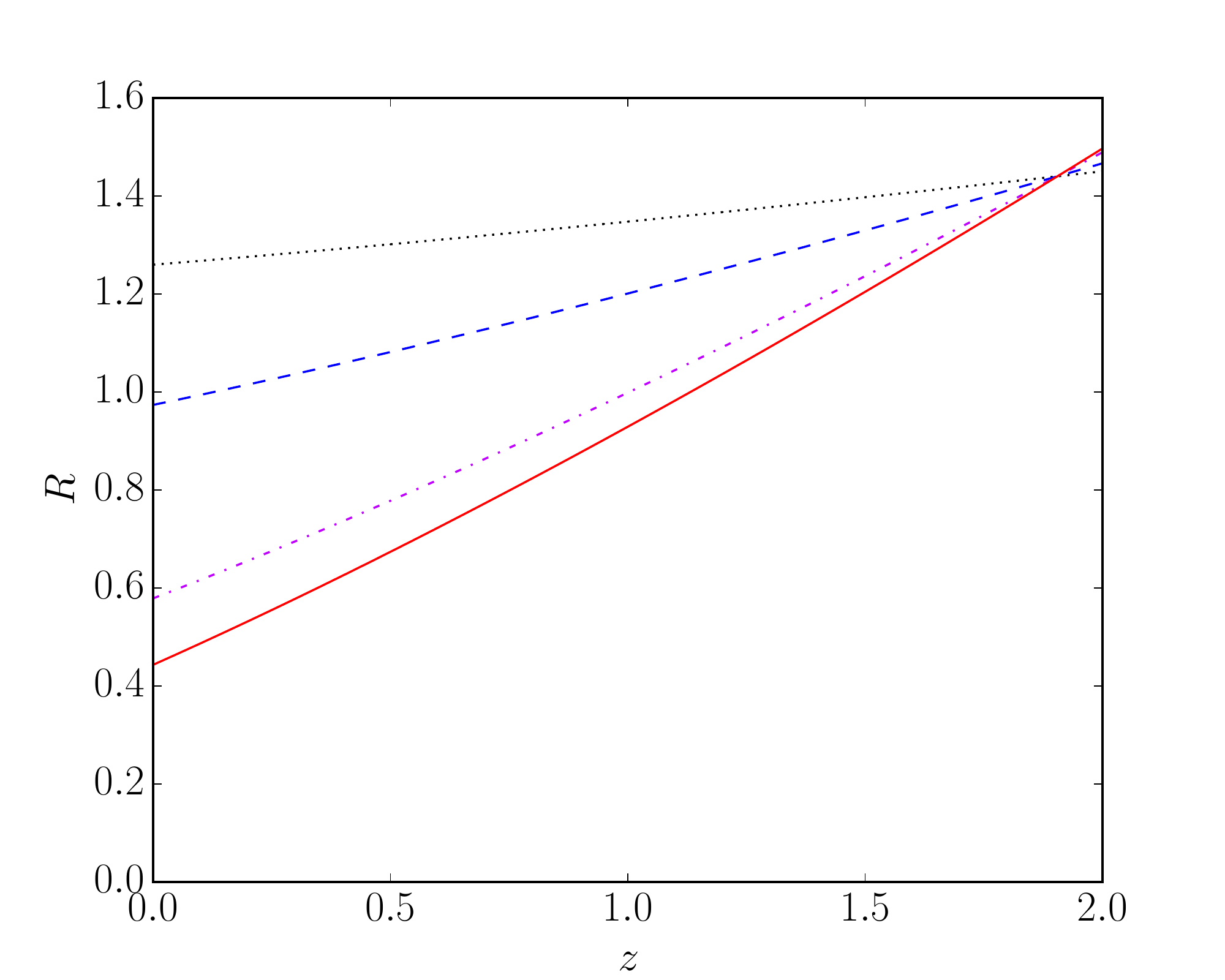}
\caption{These curves give an indication of the evolution of $R$ with
  $z$ using the model described by equation~\ref{eq:nhrefl} with the
  $N_{\mathrm{H}}$ distribution given by \citep{burlon11} and $f_2$
  given by equation~\ref{eq:type2} with $\xi = 1.3$. $f_{\mathrm{CT}}$
  is held fixed at the $z=0$ value. The evolution is shown at
  $L_{\mathrm{X}} = 10^{43}$~erg~s$^{-1}$ (dotted line),
  $L_{\mathrm{X}} = 10^{43.5}$~erg~s$^{-1}$~(dashed line),
  $L_{\mathrm{X}} = 10^{44}$~erg~s$^{-1}$~(dash-dot line), and
  $L_{\mathrm{X}} = 10^{45}$~erg~s$^{-1}$~(solid line).} 
\label{fig:avgrofz}
\end{figure}

The parameters describing the $R(N_{\mathrm{H}})$ model are
constrained by the requirement to fit the \bat\ number counts data and
to simultaneously reach the \nustar\ number counts. The model shown in
Fig.~\ref{fig:countsv2} has
$R_{\mathrm{min}}=0.1$. Such a low value is required to ensure the fit
to the \bat\ counts at high fluxes; indeed if $R_{\mathrm{min}} \ga
0.2$ then the fit to those data points is lost. At the high
$N_{\mathrm{H}}$ end, $R_{\mathrm{max}}$ is set to $1.7$ to bring the
model close to the \nustar\ data. Again, if $R_{\mathrm{max}}$ is
increased above $\approx 2$ then the model is pulled away from the \bat\ data
at high fluxes. Finally, we find
$N_{\mathrm{H,mid}}$=10$^{23}$~cm$^{-2}$ as even a value of
$10^{22.5}$~cm$^{-2}$ will lead to too strong a reflection signal at
high fluxes. To illustrate the $R$ values that result from these
parameters Fig.~\ref{fig:avgrofz} shows the $z$ evolution of the
average reflection fraction at four different AGN luminosities, where
the average is over the $N_{\mathrm{H}}$ distribution (defined by
\citealt{burlon11}) and incorporates the correct $f_2$ at the given
$L_{\mathrm{X}}$ and $z$ (Eq.~\ref{eq:type2}). 
The shape of the curves demonstrate the impact of the various factors
described above. At low redshift, more luminous AGNs, which are less
obscured than lower luminosity AGNs, have \emph{on average}
consistently weaker values of $R$, as described by eqs.~\ref{eq:type2}
and~\ref{eq:nhrefl}. As redshift increases, the obscured fraction of
AGNs with $L_{\mathrm{X}} \ga 10^{43}$~erg~s$^{-1}$ grows, and, according to
eq.~\ref{eq:nhrefl}, more obscured AGNs lead to a larger average
$R$. The shape of the $f_2$ relation (eq.~\ref{eq:type2}) leads to a
faster increase in the average $R$ for high luminosity AGNs than lower
luminosity sources. Indeed, in this model, all AGNs have an average
$R$ of $\approx 1.3$ by $z=2$.   

\begin{figure}
\includegraphics[angle=-90,width=0.5\textwidth]{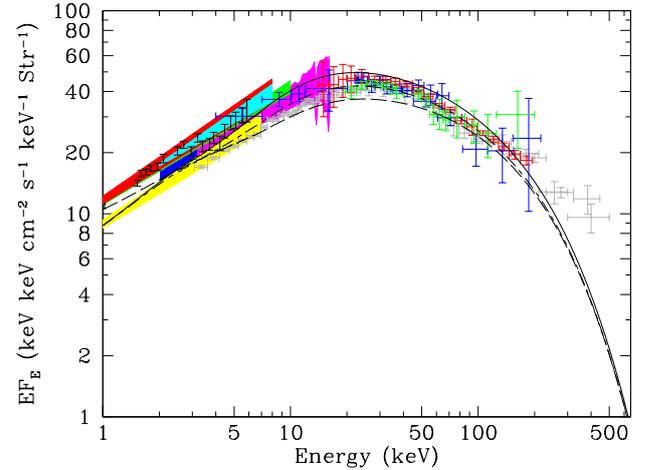}
\caption{As in Fig.~\ref{fig:cxrb}, but the curves now show models
  that are based on the $R$-$N_{\mathrm{H}}$ and the
  $f_{\mathrm{CT}}(z)$ relations (eqs.~\ref{eq:nhrefl}
  and~\ref{eq:ct}), with $\gamma=1.7$ (solid line) and $0$
  (short-dash-long-dash line). The long-dashed line plots is taken
  from Fig.~\ref{fig:cxrb} and shows a model with no redshift
  evolution in $R$ or $f_{\mathrm{CT}}$.} 
\label{fig:cxrbv2}
\end{figure}

\begin{figure}
\includegraphics[angle=-90,width=0.48\textwidth]{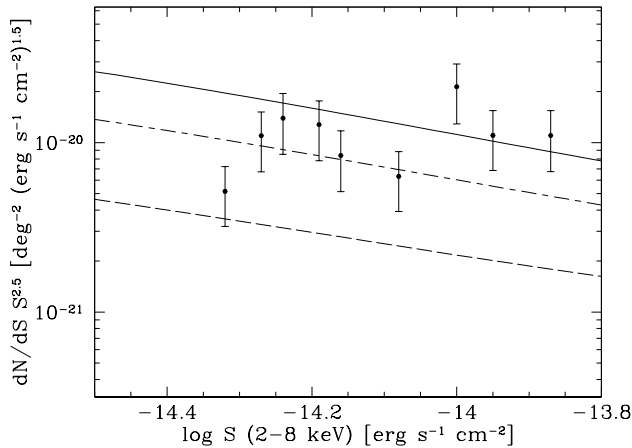}
\caption{\label{fig:ctv2} The solid line is the predicted
  Euclidean-normalized differential number counts of Compton-thick
  AGNs in the $2$--$8$~\kev\ band from the $R(N_{\mathrm{H}})$ model
  described in the text. In this model, the Compton-thick AGN fraction
  grows with redshift as described by Eq.~\ref{eq:ct} with
  $\gamma=0.7$. Alternatively, the short-dash-long-dash line plots the
  same model with no increase in the Compton-thick fraction (i.e.,
  $\gamma=0$). The dashed line and the data points are the same as in
  Fig.~\ref{fig:comptonthick}. As before, we neglect the very small
  correction between the $2$--$7$~keV and the $2$--$8$~\kev\ energy
  bands.} 
\end{figure}

The short-dash-long-dash line in Figure~\ref{fig:cxrbv2} shows that the integrated XRB spectrum
predicted by this model provides a good description of the observed
XRB. This contrasts with the previous model from
Sect.~\ref{sect:toymodel} with no redshift evolution (long dashed line
in Fig.~\ref{fig:cxrb}), and indicates that the magntidue of redshift
evolution of $R$ predicted by Eq.~\ref{eq:nhrefl} is
reasonable. However, the growth in $R$ driven by the redshift evolution of the obscured AGN
fraction is not enough on its own to account for the
\nustar\ $8$--$24$~\kev\ number counts.
Therefore, it is interesting to consider if redshift evolution
of the Compton-thick fraction may provide a better fit to the \nustar\
number counts. The solid lines in Figs.~\ref{fig:countsv2}
and~\ref{fig:cxrbv2} shows that the $R$-$N_{\mathrm{H}}$ connection
described by eq.~\ref{eq:nhrefl} combined with an evolving
Compton-thick fraction with $\gamma=0.7$ does improve the
description of the $8$--$24$~\kev\ number counts observed by
\nustar\ and \bat.  Higher values of
$\gamma$ are ruled out as they result in overpredicting the peak of
the XRB spectrum\footnote{This also implies that a model that
  only included evolution of the Compton-thick fraction (and no
  increase of $R$ with $z$) would be unable to simultaneously fit the
  $8$--$24$~\kev\ number counts and the XRB spectrum.}. A value of $\gamma=0.7$ implies a Compton-thick
fraction at $z=2$ that is $2.2\times$ larger than at $z=0$. The
predicted $2$--$8$~\kev\ number counts of Compton-thick AGNs from both
this
model and the $\gamma=0$ model are compared to the COSMOS Legacy data in Fig.~\ref{fig:ctv2}
(solid line).   
In contrast to the models presented in Sect.~\ref{sect:toymodel}
(e.g., the dashed line), the Compton-thick number counts predicted by
both models provide a decent description of the COSMOS data. Indeed,
the Compton-thick fractions at a $8$--$24$~\kev\ flux
of $2.7\times 10^{-14}$~erg~cm$^{-2}$~s$^{-1}$ are $0.11$ ($\gamma=0$)
and $0.15$ ($\gamma=0.7$). The $\gamma=0$ value is in excellent agreement
with the \nustar-derived value found by \citep{masini18}. The increase in the
Compton-thick number counts and fraction in the $R(N_{\mathrm{H}})$ model with no
Compton-thick evolution ($\gamma=0$; the
short-dash-long-dash line) is a result of the larger $R$
at faint fluxes\footnote{Recall that AGN spectra with $\log
  (N_{\mathrm{H}}/\mathrm{cm}^{-2}) =24$ and $24.5$ are constructed by
  suppressing 'standard' AGN spectra that include a reflection
  component. Therefore, the model Compton-thick spectra are influenced
  by the assumed reflection strength.}. Although this model can
describe the COSMOS data and the XRB spectrum, Fig.~\ref{fig:countsv2}
shows that a low-$\gamma$ model would not be able to match the
\nustar\ $8$--$24$~\kev\ number counts, requiring larger values of $R$
to compensate which, as described above, would lose the fit to the
\bat\ number counts. We are forced to conclude that some details of
the $R(N_{\mathrm{H}})$ model (eq.~\ref{eq:nhrefl}), which assumes the
\citet{burlon11} $N_{\mathrm{H}}$ distribution and the
$f_2(L_{\mathrm{X}})$ relationship (eq.~\ref{eq:type2}), must be
revised to self-consistently fit all the data. Nevertheless, it is
clear that a XRB model that connects the reflection strength to the
changing gas environment around AGNs can successfully describe the
hard X-ray survey data produced by \nustar\ and \bat. 

\section{Discussion}
\label{sect:discuss}
The results of this paper show that strong redshift evolution in the
average AGN reflection fraction appears to be necessary to
simultaneously describe the \nustar\ and \bat\ $8$--$24$~\kev\ number
counts. This effect was not needed in earlier XRB models as survey
data at energies $< 10$~\kev\ are not very sensitive to changes to $R$,
even at very faint flux levels (see Appendix~A). The inclusion of a
redshift evolution in one of the key XRB parameters presents a
challenge to all future XRB synthesis models, as there are many
choices on how to parameterize the evolution, and it is likely that
other parts of the problem (e.g., the $N_{\mathrm{H}}$ distribution,
the Compton-thick fraction) will also vary with $z$. Incorporating all
available survey data into a XRB model fit \citep[e.g.,][]{ananna18}
will be helpful, but may be insufficient without significant improvement
in survey data at energies $> 10$~\kev.  

The implications of an evolving $R$ are also significant for our
understanding of the changes ongoing within the AGN environment. A
changing $R$ means that the amount of high column density gas (i.e.,
gas with $\log (N_{\mathrm{H}}/\mathrm{cm}^{-2}) \geq 23.5$) is
evolving over time due to processes within the nuclear
environments. The model presented in the previous section required
strong reflection only for AGNs with the highest obscuration, as
suggested by recent observations \citep[e.g.,][]{ew16,lanz18}, in
complete contrast with the expectations of the orientation-based
unification model. Therefore, these highly obscured AGNs are evolving
separately from the more weakly obscured and unobscured AGNs and may be
connected to changes to the Compton-thick population. When combined with the overall increase in
obscured AGNs with redshift \citep[e.g.,][]{merloni14,liu17}, these
considerations all suggest a rich and complex interplay between AGN
accretion physics, the obscuration environment, and processes within
the host galaxies \citep[e.g.,][]{db10,dale15,ricci17}. Indeed, as both
the star-formation rate density and black hole accretion rate density
evolve rapidly from $z=0$ to $\approx 2$ \citep[e.g.,][]{md14}, it is perhaps not
surprising that other aspects of the AGN environment demonstrate
redshift evolution. Therefore, XRB
synthesis modeling may need to start including the predictions of
physical models of AGN obscuration and its evolution in order to
account for all these various effects. Such an exercise could be an
important way of discriminating among different AGN environment and
evolution models. 

An example of this approach was performed by \citet{gb18} who
considered nuclear starburst discs (NSDs) as the source of the
obscuring gas around AGNs. These authors used models of star-forming
discs at scales of $\sim 1$--$100$~pc from the SMBH
\citep{tqm05,ball08,gb17}, plus observations of the redshift
dependence of the gas fraction of galaxies \citep[e.g.,][]{tacc13}, to
predict the evolving $N_{\mathrm{H}}$ distribution of the obscuring
gas. Without any tuning of parameters, the model predicted that $f_2$
and $f_{\mathrm{CT}}$ should increase from $z=0$ to $2$ as
$(1+z)^{1.2}$ and $(1+z)^{1.45}$, respectively. These dependencies on
redshift are interestingly close to the ones ($1.3$ and $0.7$) needed by
the XRB synthesis model in the previous section. However, when
self-consistently including the reflection strength based on the
evolving $N_{\mathrm{H}}$ distribution, \citet{gb18} found that the
NSD model is not able to entirely explain the XRB spectrum, but
requires a large fraction of obscured high luminosity AGNs to fit the
data. Indeed, the physics of the NSD model limits its applicability to
Seyfert-like AGN luminosities \citep{ball08}, but it is an interesting
first step on a possible physical approach to XRB modeling. 

Compton-thick AGNs have played an important, but poorly understood,
role in modeling the XRB. The difficulty in detecting and
characterizing these heavily obscured AGNs is well known
\citep[e.g.,][]{ha18}, especially outside the local Universe. Advances
in modeling the X-ray spectra of deeply embedded AGNs
\citep[e.g.,][]{my09,bn11,balo18}, as well as \nustar\ observations
\citep[e.g.,][]{balo14,ann15,boor16,ann17}, have allowed progress in
identifying Compton-thick sources, but precise measurements of their
population statistics remains sparse. Figure~\ref{fig:comptonthick}
demonstrates that a fixed Compton-thick fraction, normalized to the
local space density measured by \citet{buch15}, can not match the
number counts of faint Compton-thick AGNs characterized in the COSMOS
Legacy survey, implying that either $f_{\mathrm{CT}}$ or $R$ must evolve in some
way. In addition, the previous section found that adding a simple
redshift evolution of $f_{\mathrm{CT}}$ (eq.~\ref{eq:ct}) to the
$R(N_{\mathrm{H}})$ prescription allowed for the best description of
the \nustar\ $8$--$24$~\kev\ number counts. It is possible that the
evolution of $R$ and $f_{\mathrm{CT}}$ are physically connected,
especially as large $R$ values are associated with significant
covering factors of Compton-thick gas. These results support the idea
that heavily obscured AGNs may be more commonly associated with
specific events in galaxy evolution that funnel large amounts of gas
towards the nucleus (e.g., merger events; \citealt{db10,dale15}). 

The approach taken in this paper is to focus on models that can
describe the XRB spectrum and $8$--$24$~\kev\ number counts. As seen
in Appendix~A, changes to the evolution of $R$ and Compton-thick
fraction, have a modest impact on the $2$--$10$~\kev\ counts. In
addition, we have employed the \citet{ueda14} HXLF throughout the
calculations, and considered the impact of allowing the parameters
describing AGN spectra to vary with redshift. An alternative approach,
recently pursued by \citet{ananna18}, considers a small number of
fixed AGN spectral models, but modifies the HXLF in order to fit the
X-ray survey data. While it is important that the measured HXLFs be
continuously improved, the observed increase in $f_2$
\citep[e.g.,][]{liu17}, the connection between $N_{\mathrm{H}}$ and
$R$ \citep[e.g.,][]{llanz18}, and the increasing evidence for
fundamental physical connections between the AGN and its environment
\citep[e.g.,][]{ricci17}, all strongly suggest that the observed AGN
X-ray SED will be functions of both redshift and luminosity that
should be considered in future XRB synthesis models.  

\section{Conclusions}
\label{sect:concl}
Since their advent in the mid-1990s, XRB synthesis modeling has been
an important component in the study of the demographics and evolution
of AGNs. The results of this paper, which presents evidence that the
reflection fraction $R$ evolves with $z$, implies that XRB synthesis
modeling, when combined with X-ray surveys at energies $> 10$~\kev,
should now be considered as a method to explore the evolution of AGN
\emph{physics} in addition to their demographics. The dependence of
$R$ with physical properties such as the $N_{\mathrm{H}}$ distribution
and the AGN luminosity means that the evolution of $R$ can help
distinguish between different physical models of the origin of the
obscuring gas and its connection to processes in the AGN host
galaxy. As both the photon-index and high-energy cutoff of the AGN
power-law also depend on the fundamental physics of accretion discs,
future XRB synthesis models have the potential to reveal the evolution
of many aspects of AGN physics throughout cosmic time.  

The results of this paper also provide a striking illustration of the
potential of future hard X-ray surveys in understanding the evolution
of the physical environment of AGNs. Only surveys at X-ray energies $>
10$~\kev\ will be sensitive enough to the effects of evolution in $R$
and $E_{\mathrm{cut}}$ to constrain models of AGN evolution in a
rigorous way. The hard X-ray band is also crucial to properly model
the complex SEDs of Compton-thick AGNs
\citep[e.g.,][]{balo18}. Therefore, future X-ray mission concepts that
include hard X-ray capabilities (e.g., \textit{HEX-P},
\textit{STROBE-X}; \citealt{ray18}) will be crucial in allowing XRB synthesis modeling
to reach its potential.    

\section*{Acknowledgments}

The authors thank E.\ Hollingworth for help at the outset of the
project, and J.\ Aird for providing the \nustar\ and \bat\ number
counts data. MSAM was supported in part by a Georgia Tech President's
Undergraduate Research Salary Award. 

%%%%%%%%%%%%%%%%%%%%%%%%%%%%%%%%%%%%%%%%%%%%%%%%%%

%%%%%%%%%%%%%%%%%%%% REFERENCES %%%%%%%%%%%%%%%%%%
% The reference list should include no bold or italic, no commas after
% author surnames, and no ampersand between the final two author
% names. List all of the authors if there are eight or fewer,
% otherwise give just the first author followed by ‘et al.’. 

% Author year journal volume page

\appendix
\section{The View in the 2--10 keV Band}
\label{app:a1}
\begin{figure}
\includegraphics[angle=-90,width=0.5\textwidth]{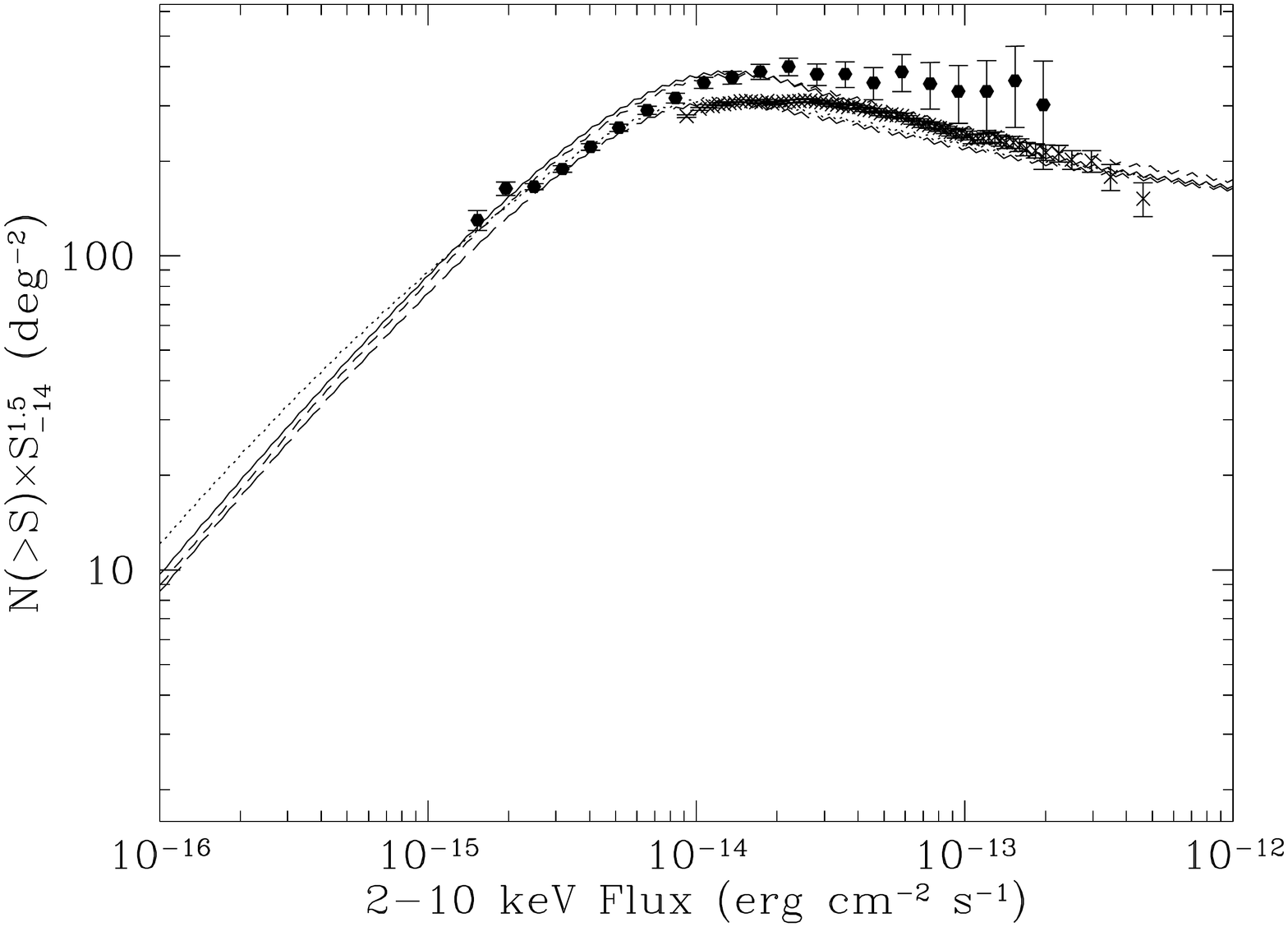}
\caption{The Euclidean-normalized integrated AGN number counts in the $2$--$10$~\kev\ band
  predicted by the four XRB synthesis models described in
  Sect.~\ref{sect:toymodel} and in Fig.~\ref{fig:counts}. The solid points are
  the counts measured by the \chandra-COSMOS survey \citep{civano16}
  while the higher-flux data are from the \xmm\ measurements of
  \citet{mat08}. These two datasets capture the spread of the
  number counts meaurements at intermediate fluxes \citep{civano16}. Despite the
  differences in the value and evolution of $R$, the different XRB
  models are all consistent with the current measurements. Therefore,
  without a significant decrease in the spread of the number count measurements at $S \approx
  10^{-14}$~erg~cm$^{-2}$~s$^{-1}$ the
  $2$--$10$~\kev\ band is not sensitive enough to
  constrain or test for significant variations of $R$ with either
  $L_{\mathrm{X}}$ or $z$. } 
\label{fig:2to10}
\end{figure}

The AGN number counts in the $2$--$10$~\kev\ energy band have been
probed to remarkable depths by both \xmm\ and
\chandra\ \citep[e.g.,][]{liu17}, but these data are not highly sensitive to
the high-energy properties of AGN spectra even at very faint
fluxes. This is illustrated in Figure~\ref{fig:2to10} which plots the
Euclidean-normalized integrated $N(>S)$ $2$--$10$~\kev\ number counts predicted by the four
models described in Sect.~\ref{sect:toymodel}. 
The four models provide a good description of number counts
measurements from \chandra\ and \xmm\ \citep{mat08,civano16}. These
datasets illustrate the spread of number counts measurements found
from a number of surveys \citep{civano16}. The four models are virtually
indistinguishable at $S \ga 10^{-13}$~erg~cm$^{-2}$~s$^{-1}$. At faint
fluxes ($S \la 10^{-15}$~erg~cm$^{-2}$~s$^{-1}$)  the
model with the boosted Compton-thick fraction (dotted line;
Sect.~\ref{sub:comptonthick}) begins to seperate from the
other models, but characterizing Compton-thick AGNs at such faint
fluxes is a significant challenge. At intermediate fluxes (e.g., $S
\approx 10^{-14}$~erg~cm$^{-2}$~s$^{-1}$) the models seperate into two
groups: the  number counts predicted by the fixed-$R$
(short-dashed line) and the strongly-evolving $R$ (solid line)
models trace the upper end of the data envelope, while the low-$R$
models (long dashed and dotted
lines) track the lower range. Therefore, use of the $2$--$10$~\kev\
band to distinguish between different luminosity and $z$ evolutions of
$R$ would require a very significant decrease in the spread of the
number count measurements.
This result emphasizes the importance of
sensitive hard X-ray measurements by \nustar\ and other future
missions, as only surveys at energies $\ga 10$~\kev\ will more easily be able to
detect aspects of AGN evolution connected to reflection and the
high-energy cutoff. 
%%%%%%%%%%%%%%%%%%%%%%%%%%%%%%%%%%%%%%%%%%%%%%%%%%

% Don't change these lines
\bsp	% typesetting comment
\label{lastpage}

\begin{thebibliography}{99}
\bibitem[\protect\citeauthoryear{Aird \etal}{2015}]{aird15} Aird J., \etal, 2015, \apj, 815, 66
\bibitem[\protect\citeauthoryear{Ajello \etal}{2008}]{ajello08} Ajello M., \etal, 2008, \apj, 689, 666
\bibitem[\protect\citeauthoryear{Ajello \etal}{2012}]{ajello12} Ajello M., Alexander D.M., Greiner J., Madejski G.M., Gehrels N., Burlon, D., 2012, \apj, 749, 21
\bibitem[\protect\citeauthoryear{Ananna \etal}{2019}]{ananna18} Ananna
  T.T., \etal, 2019, \apj, 871, 240
\bibitem[\protect\citeauthoryear{Annuar \etal}{2015}]{ann15} Annuar A., \etal, 2015, \apj, 815, 36
\bibitem[\protect\citeauthoryear{Annuar \etal}{2017}]{ann17} Annuar A., \etal, 2017, \apj, 836, 165
\bibitem[\protect\citeauthoryear{Arnaud}{1996}]{arn96} Arnaud K.A., 1996, in Jacoby G., Barnes J., eds, Astronomical Data Analysis Software and Systems V, ASP Conf.\ Ser.\ Vol.\ 101, 17
\bibitem[\protect\citeauthoryear{Akylas \&
    Georgantopoulos}{2019}]{ag19} Akylas A., Georgantopoulos I., 2019,
  \aap, in press (arXiv:1902.05137)
\bibitem[\protect\citeauthoryear{Akylas \etal}{2012}]{aky12} Akylas A., Georgakakis A., Georgantopoulos I., Brightman M., Nandra, K., 2012, \aap, 546, A98
\bibitem[\protect\citeauthoryear{Balokovi\'{c} \etal}{2014}]{balo14} Balokovi\'{c} M., \etal, 2014, \apj, 794, 111
\bibitem[\protect\citeauthoryear{Balokovi\'{c} \etal}{2018}]{balo18} Balokovi\'{c} M., \etal, 2018, \apj, 854, 42
\bibitem[\protect\citeauthoryear{Ballantyne, Everett \& Murray}{Ballantyne \etal}{2006}]{bem06} Ballantyne D.R., Everett J.E., Murray, N., 2006, \apj, 639, 740
\bibitem[\protect\citeauthoryear{Ballantyne \etal}{2011}]{ball11} Ballantyne D.R., Draper A.R., Madsen K., Rigby J.R., Treister, E., 2011, \apj, 736, 56
\bibitem[\protect\citeauthoryear{Ballantyne}{2008}]{ball08} Ballantyne D.R., 2008, \apj, 685, 787
\bibitem[\protect\citeauthoryear{Ballantyne}{2014}]{ball14} Ballantyne D.R., 2014, \mnras, 437, 2845
\bibitem[\protect\citeauthoryear{Bhayani \& Nandra}{2011}]{bn11} Bhayani S., Nandra, K., 2011, \mnras, 416, 629
\bibitem[\protect\citeauthoryear{Boorman \etal}{2016}]{boor16} Boorman P.G. \etal, 2016, \apj, 833, 245
\bibitem[\protect\citeauthoryear{Brightman \& Nandra}{2011}]{bright11} Brightman M., Nandra K., 2011, \mnras, 413, 1206
\bibitem[\protect\citeauthoryear{Buchner \etal}{2015}]{buch15} Buchner J., \etal, 2015, \apj, 802, 89
\bibitem[\protect\citeauthoryear{Burlon \etal}{2011}]{burlon11} Burlon D., Ajello M., Greiner J., Comastri A., Merloni A., Gehrels N., 2011, \apj, 728, 58
\bibitem[\protect\citeauthoryear{Cappelluti \etal}{2017}]{capp17} Cappelluti N., \etal, 2017, \apj, 837, 19
\bibitem[\protect\citeauthoryear{Churazov \etal}{2007}]{chur07} Churazov E. \etal, 2007, \aap, 467, 529
\bibitem[\protect\citeauthoryear{Civano \etal}{2015}]{civano15} Civano F., \etal, 2015, \apj, 808, 185
\bibitem[\protect\citeauthoryear{Civano \etal}{2016}]{civano16} Civano F., \etal, 2016, \apj, 819, 62
\bibitem[\protect\citeauthoryear{Comastri \etal}{1995}]{com95} Comastri A., Setti G., Zamorani G., Hasinger, G., 1995, \aap, 296, 1
\bibitem[\protect\citeauthoryear{de la Calle P\'{e}rez \etal}{2010}]{fero10} de la Calle P\'{e}rez I. \etal, 2010, \aap, 524, A50
\bibitem[\protect\citeauthoryear{Del Moro \etal}{2017}]{delmoro17} Del Moro A., \etal, 2017, \apj, 849, 57
\bibitem[\protect\citeauthoryear{De Luca \& Molendi}{2004}]{dm04} De Luca A., Molendi, S., 2004, \aap, 419, 837
\bibitem[\protect\citeauthoryear{Draper \& Ballantyne}{2009}]{db09} Draper A.R., Ballantyne D.R., 2009, \apj, 707, 778
\bibitem[\protect\citeauthoryear{Draper \& Ballantyne}{2010}]{db10} Draper A.R., Ballantyne D.R., 2010, \apjl, 715, L99
\bibitem[\protect\citeauthoryear{Esposito \& Walter}{2016}]{ew16} Esposito V., Walter R., 2016, \aap, 590, A49
\bibitem[\protect\citeauthoryear{Galeev, Rosner \& Vaiana}{Galeev \etal}{1979}]{grv79} Galeev A.A., Rosner, R., Vaiana G.S., 1979, \apj, 229, 318
\bibitem[\protect\citeauthoryear{Garc\'{\i}a \& Kallman}{2010}]{gk10} Garc\'{\i}a J., Kallman T.R., 2010, \apj, 718, 695
\bibitem[\protect\citeauthoryear{Gendreau \etal}{1995}]{gen95} Gendreau K.C. \etal, 1995, \pasj, 47, L5
\bibitem[\protect\citeauthoryear{Georgakakis \etal}{2017}]{georg17} Georgakakis A. \etal, 2017, \mnras,
  469, 3232
\bibitem[\protect\citeauthoryear{George \& Fabian}{1991}]{gf91} George I.M., Fabian A.C., 1991, \mnras, 249, 352
\bibitem[\protect\citeauthoryear{Gilli, Comastri \& Hasinger}{Gilli \etal}{2007}]{gch07} Gilli R., Comastri A., Hasinger G., 2007, \aap, 463, 79
\bibitem[\protect\citeauthoryear{Gohil \& Ballantyne}{2017}]{gb17} Gohil R., Ballantyne D.R., 2017, \mnras, 468, 4944
\bibitem[\protect\citeauthoryear{Gohil \& Ballantyne}{2018}]{gb18} Gohil R., Ballantyne D.R., 2018, \mnras, 475, 3543
\bibitem[\protect\citeauthoryear{Gruber \etal}{1999}]{gru99} Gruber D.E., Matteson J.L., Peterson L.E.,  Jung, G.V., 1999, \apj, 520, 124
\bibitem[\protect\citeauthoryear{Haardt \& Maraschi}{1991}]{hm91} Haardt F., Maraschi L., 1991, \apj, 380, L51
\bibitem[\protect\citeauthoryear{Haardt \& Maraschi}{1993}]{hm93} Haardt F., Maraschi L., 1993, \apj, 413, 507
\bibitem[\protect\citeauthoryear{Haardt, Maraschi \& Ghisellini}{Haardt \etal}{1994}]{hmg94} Haardt F., Maraschi L., Ghisellini G., 1994, \apj, 432, L95
\bibitem[\protect\citeauthoryear{Harrison \etal}{2013}]{harr13} Harrison F.A., \etal, 2013, \apj, 770, 103
\bibitem[\protect\citeauthoryear{Harrison \etal}{2016}]{harr16} Harrison F.A. \etal, 2016, \apj, 831, 185
\bibitem[\protect\citeauthoryear{Hasinger}{2008}]{has08} Hasinger G., 2008, \aap, 490, 905
\bibitem[\protect\citeauthoryear{Hickox \& Alexander}{2018}]{ha18} Hickox R.C., Alexander D.M., 2018, \araa, 56, 625
\bibitem[\protect\citeauthoryear{Hopkins \etal}{2006}]{hopk06} Hopkins P.F., Hernquist L., Cox, T.J., Di Matteo T., Robertson B., Springel V., 2006, \apjs, 163, 1
\bibitem[\protect\citeauthoryear{Iwasawa \& Taniguchi}{1993}]{it93} Iwasawa K., Taniguchi Y., 1993, \apj, 413, L15 
\bibitem[\protect\citeauthoryear{Kara \etal}{2016}]{kara16} Kara E., Alston W.N., Fabian A.C., Cackett E.M., Uttley P., Reynolds C.S., Zoghbi A., 2016, \mnras, 462, 511
\bibitem[\protect\citeauthoryear{Kocevski \etal}{2015}]{dale15} Kocevski D., \etal, 2015, \apj, 814, 104
\bibitem[\protect\citeauthoryear{Krivonos \etal}{2010}]{kriv10} Krivonos R., Tsygankov S., Revnivtsev M., Grebenev S., Churazov E., Sunyaev R., 2010, A\&A, 523, A61
\bibitem[\protect\citeauthoryear{Kushino \etal}{2002}]{kush02} Kushino A., Ishisaki Y., Morita U., Yamasaki N.Y., Ishida M., Ohashi T., Ueda, Y., 2002, \pasj, 54, 327
\bibitem[\protect\citeauthoryear{La Franca \etal}{2005}]{laf05} La Franca F. \etal, 2005, ApJ, 635, 864
\bibitem[\protect\citeauthoryear{Lansbury \etal}{2017}]{lans17} Lansbury G., \etal, 2017, \apj, 836, 99
\bibitem[\protect\citeauthoryear{Lanz \etal}{2019}]{llanz18} Lanz L.,
  \etal, 2019, \apj, 870, 26
\bibitem[\protect\citeauthoryear{Lanzuisi \etal}{2018}]{lanz18} Lanzuisi G., \etal, 2018, \mnras, 480, 2578
\bibitem[\protect\citeauthoryear{Liu \etal}{2017}]{liu17} Liu T., \etal, 2017, \apjs, 232, 8
\bibitem[\protect\citeauthoryear{Lumb \etal}{2002}]{lumb02} Lumb D.H., Warwick R.S., Page M., De Luca, A., 2002, \aap, 389, 93
\bibitem[\protect\citeauthoryear{MacLeod \etal}{2015}]{mac15} MacLeod
  C.L., \etal, 2015, \apj, 806, 258
\bibitem[\protect\citeauthoryear{Madau \& Dickinson}{2014}]{md14} Madau P.,
  Dickinson M., 2014, \araa, 52, 415
\bibitem[\protect\citeauthoryear{Mainieri \etal}{2007}]{main07} Mainieri V., \etal, 2007, \apjs, 172, 368
\bibitem[\protect\citeauthoryear{Mantovani, Nandra \& Ponti}{Mantovani \etal}{2016}]{mnp16} Mantovani G., Nandra  K., Ponti, G., 2016, \mnras, 458, 4198
\bibitem[\protect\citeauthoryear{Marchesi \etal}{2016}]{march16} Marchesi S., \etal, 2016, \apj, 830, 100
\bibitem[\protect\citeauthoryear{Masini \etal}{2018}]{masini18} Masini
  A., \etal, 2018, \apjs, 235, 17
\bibitem[\protect\citeauthoryear{Mateos \etal}{2008}]{mat08} Mateos  S., \etal, 2008, \aap, 492, 51
\bibitem[\protect\citeauthoryear{Matt, Perola \& Piro}{Matt \etal}{1991}]{matt91} Matt G., Perola G.C.,Piro, L., 1991, \aap, 247, 25
\bibitem[\protect\citeauthoryear{Merloni \etal}{2014}]{merloni14} Merloni A., \etal, 2014, \mnras, 437, 3550 
\bibitem[\protect\citeauthoryear{Moretti \etal}{2009}]{mor09} Moretti A. \etal, 2009, \aap, 493, 501
\bibitem[\protect\citeauthoryear{Morrison \& McCammon}{1983}]{mcm83} Morrison R., McCammon D., 1983, \apj, 270, 119
\bibitem[\protect\citeauthoryear{Mullaney \etal}{2015}]{mull15} Mullaney J.R., \etal, 2015, \apj, 808, 184
\bibitem[\protect\citeauthoryear{Murphy \& Yaqoob}{2009}]{my09} Murphy K.D., Yaqoob T., 2009, \mnras, 397, 1549
\bibitem[\protect\citeauthoryear{Nandra}{2006}]{nan06} Nandra K., 2006, \mnras, 368, L62
\bibitem[\protect\citeauthoryear{Nandra \etal}{2007}]{nan07} Nandra K., O'Neill P.M., George I.M., Reeves J.N., 2007, \mnras, 382, 194
\bibitem[\protect\citeauthoryear{Patrick \etal}{2012}]{pat12} Patrick A.R., Reeves J.N., Porquet D., Markowitz A.G., Braito V., Lobban, A.P., 2012, \mnras, 426, 2522
%\bibitem[\protect\citeauthoryear{Piccinotti \etal}{1982}]{pic82}
%Piccinotti G., Mushotzky R.F., Boldt E.A., Holt S.S., Marshall F.E.,
%Serlemitsos P.J., Shafer R.A., 1982, ApJ, 253, 485
\bibitem[\protect\citeauthoryear{Ray \etal}{2018}]{ray18} Ray P.,
  \etal, 2018, Proceedings of the SPIE, 10699, 1069919 
\bibitem[\protect\citeauthoryear{Reis \& Miller}{2013}]{rm13} Reis R.C., Miller J.M., 2013, \apj, 769, L7
\bibitem[\protect\citeauthoryear{Revnivtsev \etal}{2003}]{rev03} Revnivtsev M., Gilfanov M., Sunyaev R., Jahoda K., Markwardt C., 2003, \aap, 411, 329
\bibitem[\protect\citeauthoryear{Ricci \etal}{2011}]{ricci11} Ricci C., Walter R., Courvoisier T.J.-L., Paltani S., 2011, \aap, 532, A102
\bibitem[\protect\citeauthoryear{Ricci \etal}{2013}]{ricci13} Ricci C., Paltani S., Ueda Y., Awaki H., 2013, \mnras, 435, 1840
\bibitem[\protect\citeauthoryear{Ricci \etal}{2014}]{ricci14} Ricci C., Ueda Y., Paltani S., Ichikawa K., Gandhi P., Awaki H., 2014, \mnras, 441, 3622
\bibitem[\protect\citeauthoryear{Ricci \etal}{2015}]{ricci15} Ricci C., Ueda Y., Koss M.J., Trakhtenbrot B., Bauer F.E., Gandhi P., 2015, \apjl, 815, 13
\bibitem[\protect\citeauthoryear{Ricci \etal}{2017}]{ricci17} Ricci C., \etal, 2017, \apjs, 233, 17
\bibitem[\protect\citeauthoryear{Ricci \etal}{2018}]{ricci18} Ricci C., \etal, 2018, \mnras, 480, 1819
\bibitem[\protect\citeauthoryear{Ross \& Fabian}{1993}]{rf93} Ross R.R., Fabian A.C., 1993, \mnras, 261, 74
\bibitem[\protect\citeauthoryear{Ross, Fabian \& Young}{Ross \etal}{1999}]{rfy99} Ross R.R., Fabian A.C., Young A.J., 1999, \mnras, 306, 461
\bibitem[\protect\citeauthoryear{Ross \& Fabian}{2005}]{rf05} Ross R.R., Fabian A.C., 2005, \mnras, 358, 211
\bibitem[\protect\citeauthoryear{Shu, Yaqoob \& Wang}{Shu \etal}{2010}]{shu10} Shu X.W., Yaqoob T., Wang, J.X., 2010, \apjs, 187, 581
\bibitem[\protect\citeauthoryear{Shu, Yaqoob \& Wang}{Shu \etal}{2011}]{shu11} Shu X.W., Yaqoob T., Wang, J.X., 2011, \apj, 738, 147
\bibitem[\protect\citeauthoryear{Shu \etal}{2012}]{shu12} Shu X.W., Wang J.X., Yaqoob T., Jiang P., Zhou Y.Y., 2012, \apj, 744, L21
\bibitem[\protect\citeauthoryear{Tacconi \etal}{2013}]{tacc13} Tacconi L.J., \etal, 2013, \apj, 768, 74
\bibitem[\protect\citeauthoryear{Thompson, Quataert \& Murray}{2005}]{tqm05} Thompson T.A., Quataert E., Murray N., 2005, \apj, 630, 167
\bibitem[\protect\citeauthoryear{Tortosa \etal}{2018}]{tort18} Tortosa A., Bianchi S., Marinucci A., Matt G., Petrucci P.O., 2018, \aap, 614, A37
\bibitem[\protect\citeauthoryear{Treister \& Urry}{2005}]{tu05} Treister E., Urry C.M., 2005, \apj, 630, 115
\bibitem[\protect\citeauthoryear{Treister, Urry \& Virani}{Treister \etal}{2009}]{tuv09} Treister E., Urry C.M., Virani S., 2009, \apj, 696, 110
\bibitem[\protect\citeauthoryear{T\"urler \etal}{2010}]{tur10} T\"urler M., Chernyakova M., Courvoisier T. J.-L., Lubi\'nski P., Neronov A., Produit N., Walter, R., 2010, \aap, 512, A49
\bibitem[\protect\citeauthoryear{Ueda \etal}{2003}]{ueda03} Ueda Y., Akiyama M., Ohta K., Miyaji T., 2003, \apj, 598, 886
\bibitem[\protect\citeauthoryear{Ueda \etal}{2014}]{ueda14} Ueda Y., Akiyama M., Hasinger G., Miyaji T., Watson, M.G., 2014, \apj, 786, 104
\bibitem[\protect\citeauthoryear{Vasudevan, Mushotzky \& Gandhi}{Vasudevan \etal}{2013}]{vmg13} Vasudevan R.V., Mushotzky R.F., Gandhi P., 2013, \apj, 770, L37
\bibitem[\protect\citeauthoryear{Vecchi \etal}{1999}]{vec99} Vecchi A., Molendi S., Guainazzi M., Fiore F., Parmar A. 1999, \aap, 349, L73
\bibitem[\protect\citeauthoryear{Walton \etal}{2013}]{walton13} Walton D.J., Nardini E., Fabian A.C., Gallo L.C., Reis, R.C., 2013, \mnras, 428, 2901 
\bibitem[\protect\citeauthoryear{Yaqoob}{2012}]{yaq12} Yaqoob T., 2012, \mnras, 423, 3360
\bibitem[\protect\citeauthoryear{Zappacosta \etal}{2018}]{zappa18} Zappacosta L., \etal, 2018, \apj, 854, 33
\bibitem[\protect\citeauthoryear{Zoghbi \etal}{2013}]{zog13} Zoghbi A., Reynolds C., Cackett E.M., Miniutti G., Kara E., Fabian A.C., 2013, \apj, 767, 121
\end{thebibliography}
\end{document}